\documentclass[12pt]{elsarticle}
\usepackage{refmerge}
\usepackage{epsfig}
\usepackage{multicol}
\usepackage[percent]{overpic}
\usepackage{textcomp}
\usepackage{float}
\floatstyle{plaintop}
\restylefloat{table}

\begin{document}

\date{\today}

\title{\bf{STUDY OF THE PROCESS $e^+e^-{\to}K^+K^-\eta$ WITH THE CMD-3 DETECTOR AT THE VEPP-2000 COLLIDER}}

\author[adr1,adr2]{V.L.~Ivanov\fnref{tnot}}
\author[adr1,adr2]{G.V.~Fedotovich}
\author[adr1,adr2]{R.R.~Akhmetshin}
\author[adr1,adr2]{A.N.~Amirkhanov}
\author[adr1,adr2]{A.V.~Anisenkov}
\author[adr1,adr2]{V.M.~Aulchenko}
\author[adr1]{V.Sh.~Banzarov}
\author[adr1]{N.S.~Bashtovoy}
\author[adr1,adr2]{D.E.~Berkaev}
\author[adr1,adr2]{A.E.~Bondar}
\author[adr1]{A.V.~Bragin}
\author[adr1,adr2,adr3]{S.I.~Eidelman}
\author[adr1,adr2]{D.A.~Epifanov}
\author[adr1,adr2,adr4]{L.B.~Epshteyn}
\author[adr1,adr2]{A.L.~Erofeev}
\author[adr1,adr2]{S.E.~Gayazov}
\author[adr1,adr2]{A.A.~Grebenuk}
\author[adr1,adr2]{S.S.~Gribanov}
\author[adr1,adr2,adr4]{D.N.~Grigoriev}
\author[adr1,adr2]{F.V.~Ignatov}
\author[adr1]{S.V.~Karpov}
\author[adr1]{A.S.~Kasaev}
\author[adr1,adr2]{V.F.~Kazanin}
\author[adr1,adr2]{I.A.~Koop}
\author[adr1,adr2]{A.A.~Korobov}
\author[adr1,adr4]{A.N.~Kozyrev}
\author[adr1,adr2]{E.A.~Kozyrev}
\author[adr1,adr2]{P.P.~Krokovny}
\author[adr1,adr2]{A.E.~Kuzmenko}
\author[adr1,adr2]{A.S.~Kuzmin}
\author[adr1,adr2]{I.B.~Logashenko}
\author[adr1,adr2]{P.A.~Lukin}
\author[adr1]{A.P.~Lysenko}
\author[adr1]{K.Yu.~Mikhailov}
\author[adr1]{V.S.~Okhapkin}
\author[adr1,adr2]{E.A.~Perevedentsev}
\author[adr1]{Yu.N.~Pestov}
\author[adr1,adr2]{A.S.~Popov}
\author[adr1,adr2]{G.P.~Razuvaev}
\author[adr1]{A.A.~Ruban}
\author[adr1]{N.M.~Ryskulov}
\author[adr1,adr2]{A.E.~Ryzhenenkov}
\author[adr1,adr2]{A.V.~Semenov}
\author[adr1]{Yu.M.~Shatunov}
\author[adr1,adr2,adr5]{V.E.~Shebalin}
\author[adr1,adr2]{D.N.~Shemyakin}
\author[adr1,adr2]{B.A.~Shwartz}
\author[adr1,adr2]{D.B.~Shwartz}
\author[adr1,adr6]{A.L.~Sibidanov}
\author[adr1,adr2]{E.P.~Solodov}
\author[adr1]{M.V.~Timoshenko}
\author[adr1]{V.M.~Titov}
\author[adr1,adr2]{A.A.~Talyshev}
\author[adr1]{S.S.~Tolmachev}
\author[adr1]{A.I.~Vorobiov}
\author[adr1,adr2]{Yu.V.~Yudin}

\address[adr1]{Budker Institute of Nuclear Physics, SB RAS, Novosibirsk, 630090, Russia}
\address[adr2]{Novosibirsk State University, Novosibirsk, 630090, Russia}
\address[adr3]{Lebedev Physical Institute RAS, Moscow, 119333, Russia}
\address[adr4]{Novosibirsk State Technical University, Novosibirsk, 630092, Russia}
\address[adr5]{University of Hawaii, Honolulu, Hawaii 96822, USA}
\address[adr6]{University of Victoria, Victoria, BC, Canada V8W 3P6}
\fntext[tnot]{Corresponding author: V.L.Ivanov@inp.nsk.su}

\vspace{0.7cm}
\begin{abstract}
  \hspace*{\parindent}
The process $e^+e^-{\to}K^+K^-\eta$ has been studied in the center-of-mass 
energy range from 1.59 to 2.007\,GeV using the data sample of 59.5 pb$^{-1}$, 
collected with the CMD-3 detector at the VEPP-2000 $e^+e^-$ collider in 
2011, 2012 and 2017. The $K^+K^-\eta$ final state is found to be dominated
by the contribution of the $\phi(1020)\eta$ intermediate state. The cross 
section of the process $e^+e^-{\to}\phi(1020)\eta$ has been measured 
with a systematic uncertainty of 5.1$\%$ on the base of 3009 $\pm$ 67 selected events.  
The obtained cross section has been used to calculate 
the contribution to the anomalous magnetic moment of the muon:
$a_{\mu}^{\phi\eta}(E<1.8\, {\rm GeV})=(0.321 \pm 0.015_{\rm  stat} \pm 0.016_{\rm  syst}) \times 10^{-10}$, 
$a_{\mu}^{\phi\eta}(E<2.0\, {\rm GeV})=(0.440 \pm 0.015_{\rm  stat} \pm 0.022_{\rm  syst}) \times 10^{-10}$.
From the cross section approximation the $\phi(1680)$ meson parameters have been determined with better statistical precision, than in previous studies.
\end{abstract}

\maketitle

\baselineskip=17pt


\section{\boldmath Introduction}
\hspace*{\parindent}
A high-precision measurement of the cross section of $e^{+}e^{-}{\to}~hadrons$
has numerous applications including, e.g., a calculation of the hadronic 
contribution to the muon anomalous magnetic moment $(g-2)_\mu$ 
and running fine structure constant. To confirm or deny the observed 
difference between the calculated $(g-2)_\mu$ value~\cite{fred,davier,thomas,teubner}
and the measured one~\cite{bnl}, more precise measurements of the exclusive 
channels of $e^{+}e^{-}{\to}~hadrons$ are necessary.

The process $e^+e^-{\to}K^+K^-\eta$ has been previously studied by the BaBar 
collaboration at the center-of-mass energies ($E_{\rm c.m.}$) from 1.56 to 3.48 
GeV in the $\eta{\to}2\gamma$ decay mode~\cite{babar_kpkmeta_2gamma} and from 
1.56 to 2.64 GeV in the $\eta{\to}\pi^+\pi^-\pi^0$ decay 
mode~\cite{babar_kpkmeta_pippimpi0} (${\sim}480$ and ${\sim}110$ signal 
events were selected, respectively). Another study of this process in the $E_{\rm c.m.}$ range from 1.56 to 2.0 GeV 
was performed by the SND collaboration~\cite{snd_kpkmeta} with ${\sim}265$ selected signal events. 
In the BaBar study~\cite{babar_kpkmeta_2gamma} it was found that the dominant intermediate mechanism in this process is 
$e^{+}e^{-}{\to}\phi(1680){\to}\phi(1020)\eta$ (in what follows 
$\phi(1020){\equiv}\phi$, $\phi(1680){\equiv}\phi^{\prime}$ and natural units $\hbar=c=1$ are used), so the total 
cross section $\sigma(e^+e^-{\to}K^+K^-\eta)$ was subdivided into two parts: 
$\sigma(e^+e^-{\to}\phi\eta){\cdot}\mathcal{B}^{\phi}_{K^+K^-}$ (for the 
invariant masses of kaons $m_{\rm inv,\, 2K}<1045\,{\rm MeV}$) and 
$\sigma_{\rm NON-\phi}(e^+e^-{\to}K^+K^-\eta)$ 
(for $m_{\rm inv,\, 2K}>1045\,{\rm MeV}$). The latter was only 3--12\% 
of the total cross section, and the data samples of BaBar were not sufficient 
to analyze the intermediate mechanisms in the $\rm NON-\phi$ part of the 
reaction~\cite{babar_kpkmeta_2gamma}. As the $\phi^{\prime}$ meson dominates in 
this process, its parameters can be extracted from the approximation of 
the $e^+e^-{\to}\phi\eta$ cross section.

In this paper we report the results of the study of the process 
$e^+e^-{\to}K^+K^-\eta$, based on 59.5 pb$^{-1}$ of integrated luminosity 
collected by the CMD-3 detector in 2011, 2012 and 2017 in the $E_{\rm c.m.}$ 
range from 1.59 to 2.007\,GeV. We observe the contribution of the
$\phi\eta$ intermediate state only, and from the approximation of the
$e^{+}e^{-}{\to}\phi\eta$ cross section determine the parameters of 
the $\phi^{\prime}$ meson.


\section{CMD-3 detector and data set}
\hspace*{\parindent}

The VEPP-2000 $e^+e^-$ collider~\cite{vepp1,vepp2,vepp3,vepp4} at the Budker 
Institute of Nuclear Physics covers the $E_{\rm c.m.}$ range from 0.32 to 
2.01\,GeV and uses a technique of round beams to reach an instantaneous
luminosity of 10$^{32}$\,cm$^{-2}$s$^{-1}$ at $E_{\rm c.m.}$=2.0\,GeV. The 
Cryogenic Magnetic Detector (CMD-3) described in~\cite{cmd3} is installed in 
one of the two interaction regions of the collider. The detector tracking 
system consists of the cylindrical drift chamber (DC)~\cite{dc} and 
double-layer cylindrical multiwire proportional Z-chamber, installed inside a 
thin (0.085 $\rm X_{0}$) superconducting solenoid with 1.0--1.3 T magnetic 
field. Both subsystems are also used to provide the trigger signals. DC 
contains 1218 hexagonal cells in 18 layers and allows one to measure charged 
particle momentum with 1.5--4.5$\%$ accuracy in the 40--1000\,\rm MeV range, 
and the polar ($\theta$) and azimuthal ($\varphi$) angles with 20 mrad and 
3.5--8.0 mrad accuracy, respectively. Amplitude information from the DC signal 
wires is used to measure ionization losses ($dE/dx$) of charged particles. The 
barrel electromagnetic calorimeters based on liquid xenon (LXe)~\cite{lxe} 
(5.4 $\rm X_{0}$) and CsI crystals (8.1 $\rm X_{0}$) are placed outside the 
solenoid~\cite{cal}. The total amount of material in front of the barrel 
calorimeter is 0.13 $\rm X_{0}$ that includes the solenoid as well as the 
radiation shield and vacuum vessel walls. The endcap calorimeter is made of 
680 BGO crystals of 13.4 $\rm X_{0}$ thickness~\cite{cal}. The magnetic 
flux-return yoke is surrounded by scintillation counters which are used to 
tag cosmic events.

To study a detector response and determine a detection efficiency, we have 
developed a code for Monte Carlo (MC) simulation of our detector based on the 
GEANT4~\cite{GEANT4} package so that all simulated events are subjected to the 
same reconstruction and selection procedures as the data.

The energy range $E_{\rm c.m.}$ = 1.0--2.007\,GeV was scanned in the runs of 
2011, 2012 and 2017. The integrated luminosity at each energy point was 
determined using events of the processes $e^+e^-{\to}e^+e^-$ and 
$e^+e^-{\to}2\gamma$~\cite{lum}. The beam energy was 
monitored by measuring the current in the dipole magnets of the main ring 
(in 2011 and 2012), and by using the Back-Scattering-Laser-Light system 
(in 2017)~\cite{laser,laser2}. In the runs of 2011 and 2012 
we use the measured average momentum of electrons and positrons in events of 
Bhabha scattering, as well as the average momentum of proton-antiproton pairs 
from the process $e^+e^-{\to}p\bar{p}$ process~\cite{pp} to determine the 
actual $E_{\rm c.m.}$ values for each nominal beam energy with about 6 and 2\,MeV
accuracy, respectively.


\section{Study of the process $e^+e^-{\to}K^+K^-\eta$}
\hspace*{\parindent}
\subsection{Event selection\label{sec:selections}}

In order to measure the cross section of $K^{+}K^{-}\eta$ production, one needs 
to determine the detection efficiency for these events. The detection 
efficiency strongly depends on the intermediate mechanisms of the process and 
to reveal those mechanisms $K^+K^-\eta$ events are selected in the 
$\eta{\to}2\gamma$ decay mode resulting in a sample of almost background-free 
events.

\subsubsection{\label{GoodTracks} Selection of ``good"\,tracks}

Candidates for $K^+K^-\eta$ events are required to have exactly two ``good" 
tracks in the DC with the following ``good"\,track definition: 1) a track 
transverse momentum $p_{\perp}$ is larger than 60\,\rm MeV; 2) a distance 
of the closest track approach (PCA) to the beam axis in the transversal plane 
($\rho_{\rm PCA}$) is less than 0.5\,cm; 3) a distance from the PCA to the 
center of the interaction region along the beam axis ($z_{\rm PCA}$) is less 
than 12\,cm; 4) a polar angle $\theta$ of the track is in the range from 
$\theta_{\rm cut}{\equiv}0.9$ to $\pi-\theta_{\rm cut}$ radians; 5) for positively 
charged particles ionization losses $dE/dx$ of the track are smaller than 
ionization losses typical of a proton with the same momentum.


\subsubsection{\label{KPiseparation} Selection of kaons}

To select events with two oppositely charged kaons, we use the functions 
$f_{K/\pi}(p,dE/dx)$ representing the probability density for charged kaon/pion 
with the momentum $p$ to produce the energy losses $dE/dx$ in the DC. These 
functions were obtained at each $E_{\rm c.m.}$ in the analysis of the process 
$e^+e^-{\to}K^+K^-\pi^+\pi^-$ with the CMD-3 detector~\cite{shemyakin_kkpipi}, 
and we use them to simulate $dE/dx$ of the kaons and pions. 

Further, the log-likelihood function (LLF) for the hypothesis that for 
$i=1,2$ two oppositely charged particles with the momenta $p_{i}$ and energy 
losses $(dE/dx)_{i}$ are kaons is defined as

\begin{eqnarray}
  L_{\rm 2K}=\sum_{i=1}^{2}\ln\Biggl(\frac{f_{K}(p_{i},(dE/dx)_{i})}{f_{K}(p_{i},(dE/dx)_{i})+f_{\pi}(p_{i},(dE/dx)_{i})}\Biggr),
\end{eqnarray}

\noindent see its distribution in Fig.~\ref{fig:L_2K}. We apply the cut 
$L_{\rm 2K}>-0.3$ to select events with kaons.

\subsubsection{Kinematic fit}

\begin{figure}[h!]
  \begin{minipage}[t]{0.48\textwidth}
    \centerline{\includegraphics[width=0.98\textwidth]{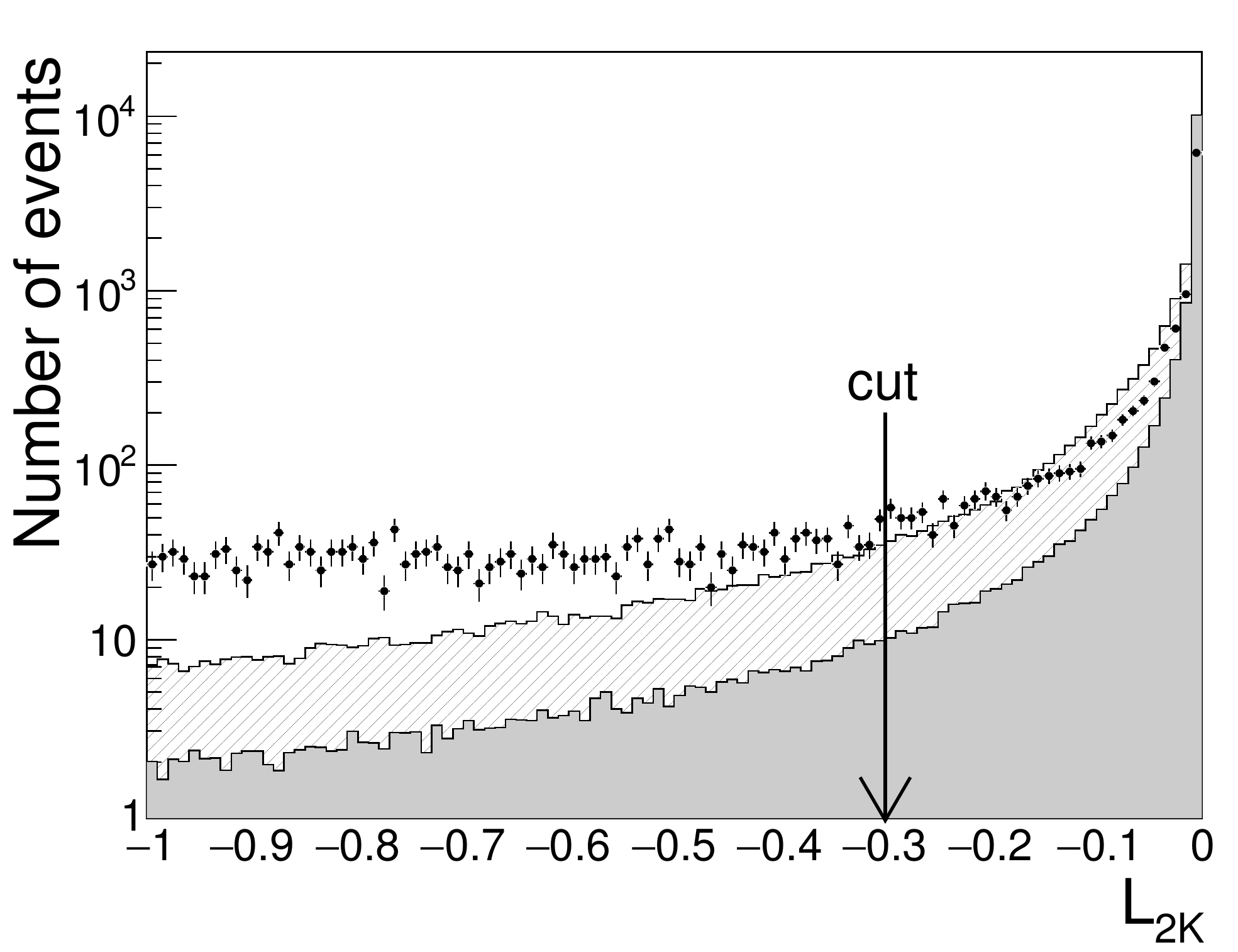}} 
    \caption{Distribution of $L_{\rm 2K}$ in data (points), simulation of 
$e^{+}e^{-}{\to}\phi\eta{\to}K^+K^-2\gamma$ (the grey histogram) and simulaion 
of $e^{+}e^{-}{\to}K^{+}K^{-}\eta{\to}K^+K^-2\gamma$ according to phase space 
(the dashed histogram). The simulated histograms are normalized to the number 
of events in the experimental one. Data at all energies are used. 
\label{fig:L_2K}}
  \end{minipage}\hfill\hfill  
  \begin{minipage}[t]{0.48\textwidth}
    \begin{overpic}[width=0.98\textwidth]{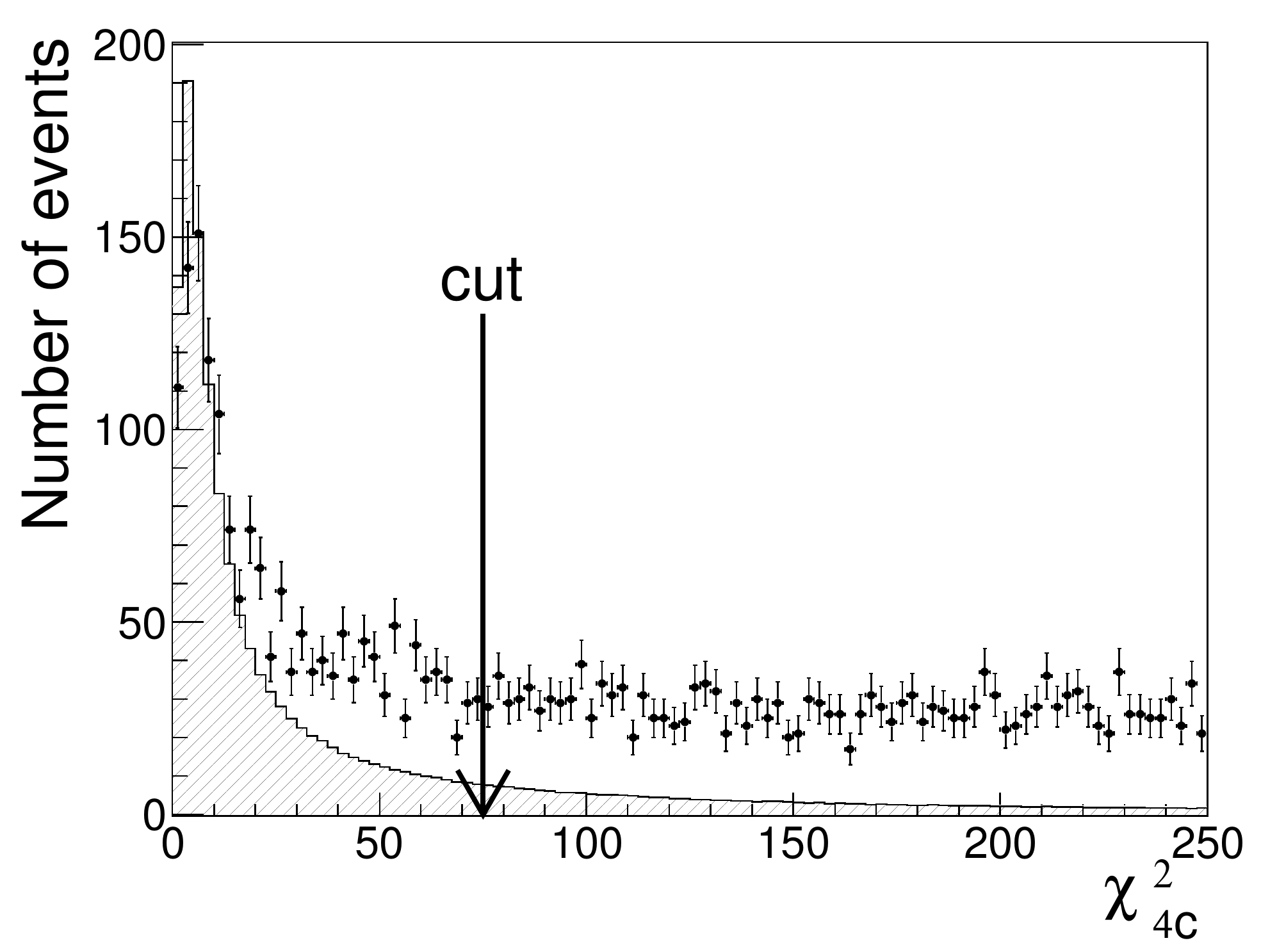}
      \put (43,28) {\includegraphics[width=0.5\textwidth]{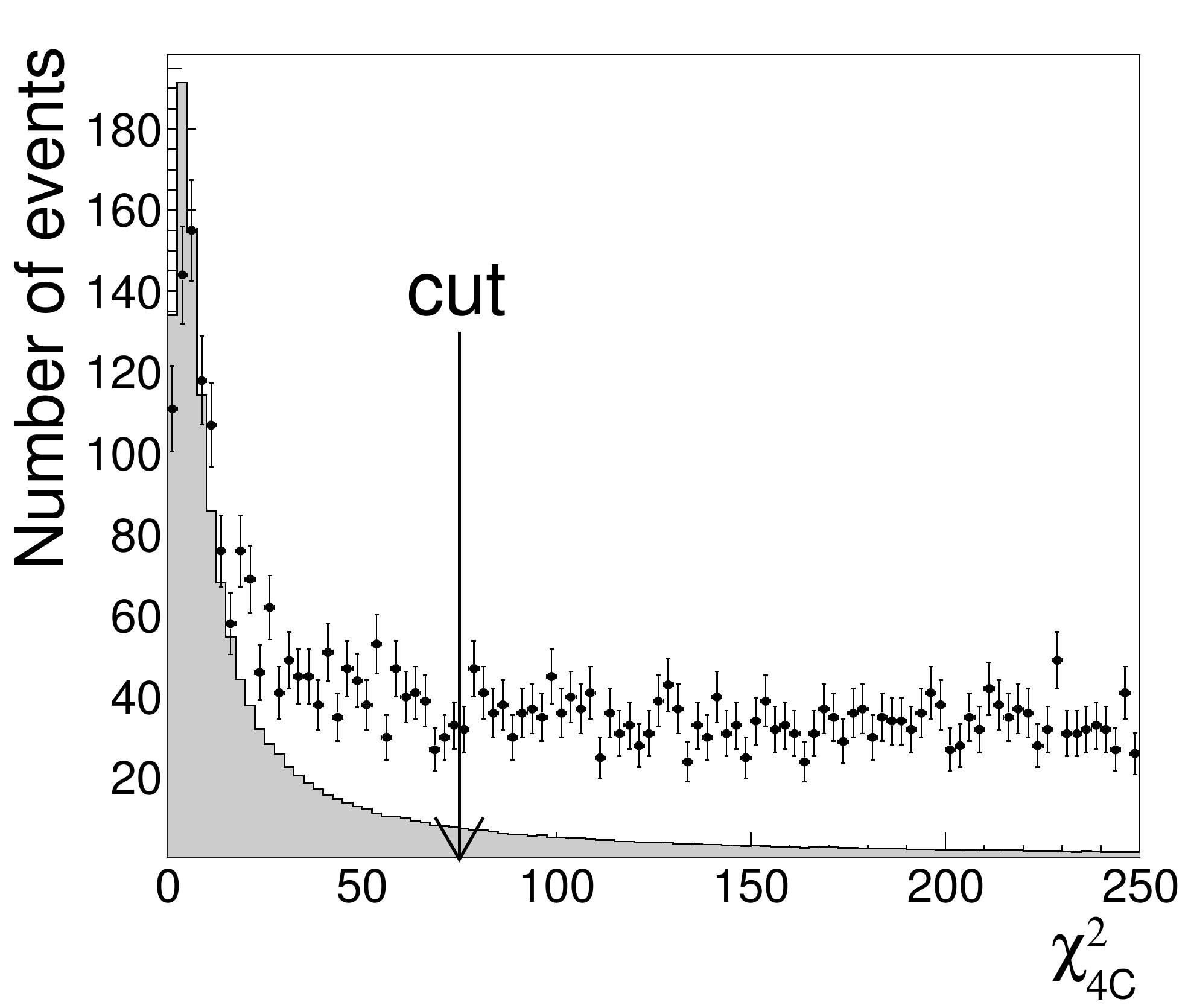}}      
    \end{overpic}
    \caption{Distribution of the $\chi^{2}_{\rm 4C}$ value in data (points) and 
simulation of $e^{+}e^{-}{\to}K^{+}K^{-}\eta{\to}K^+K^-2\gamma$ according to 
phase space (the dashed histogram), normalized to the 
$e^{+}e^{-}{\to}\phi\eta{\to}K^+K^-2\gamma$ cross section. The inset shows a
similar distribution for the simulation of 
$e^{+}e^{-}{\to}\phi\eta{\to}K^+K^-2\gamma$ (the grey histogram). Data at 
all energies are used. \label{fig:chi2}}
  \end{minipage}\hfill\hfill  
\end{figure}

To select $K^{+}K^{-}\eta$ events in the $\eta{\to}2\gamma$ mode, we select 
events with two or more photons with energies larger than 40 MeV and polar 
angles $\theta_{\rm \gamma}$ in the range from 0.5 to $\pi-0.5$ radians. Then we 
perform a kinematic fit (assuming energy-momentum conservation) of a 
$K^{+}K^{-}$ pair with each pair of selected photons, searching for the 
combination that gives the minimal $\chi^{2}_{\rm 4C}$. We apply a
requirement on the $\chi^{2}_{\rm 4C}<75$ value to select signal events, see 
Fig.~\ref{fig:chi2} (unless otherwise stated, in what follows the simulated 
histograms are normalized to the expected number of events according to the 
cross sections measured in~\cite{babar_kpkmeta_2gamma,babar_kpkmeta_pippimpi0,shemyakin_kkpipi,babar_kkpipi}; the simulation of signal and background processes
includes the emission of photon jets by initial electrons and positrons
according to~\cite{kur_fad}). 

\begin{figure}[h!]
  \begin{minipage}[t]{0.46\textwidth}
    \centerline{\includegraphics[width=0.98\textwidth]{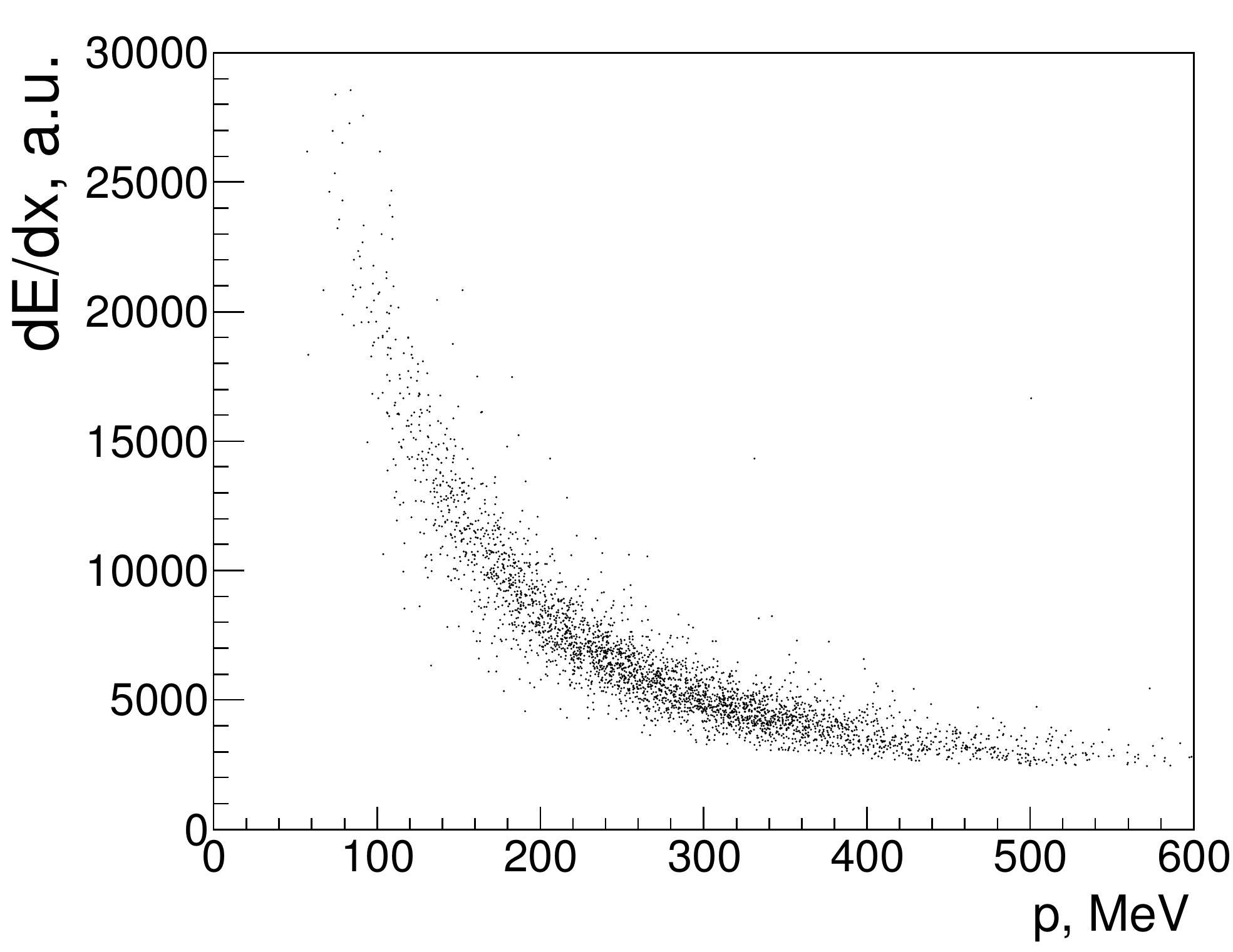}}    
    \caption{The distribution of $dE/dx$ vs particle momentum in data (all
energies are used.)\label{fig:dEdx}}
  \end{minipage}\hfill\hfill  
  \begin{minipage}[t]{0.46\textwidth}
    \centerline{\includegraphics[width=0.98\textwidth]{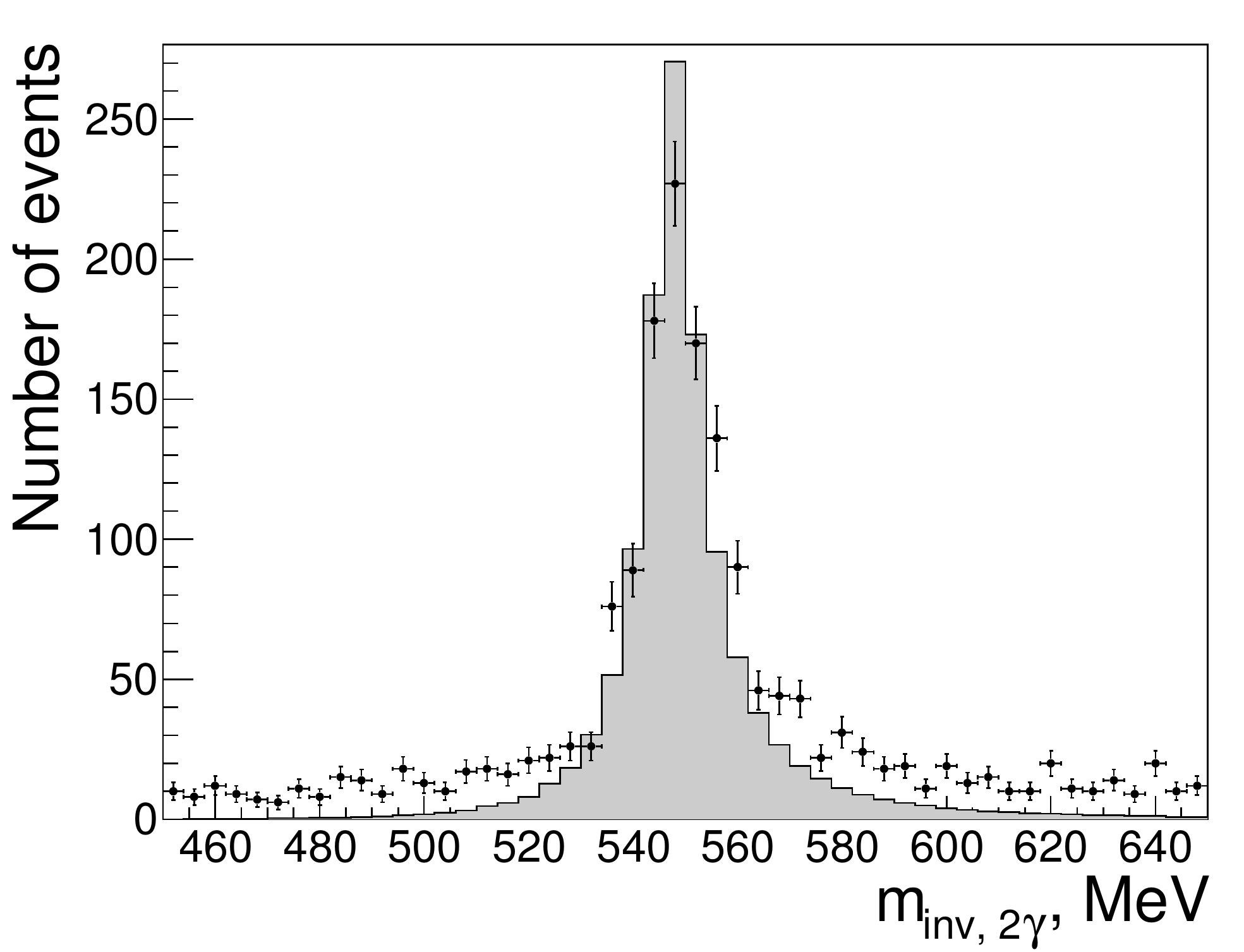}}   
    \caption{Distribution of $m_{\rm inv, 2\gamma}$ in data (points) and 
simulation of $e^{+}e^{-}{\to}\phi\eta{\to}K^+K^-2\gamma$ (the grey histogram). 
Data at all energies are used.          \label{fig:m_2g}}
  \end{minipage}\hfill\hfill  
\end{figure}

\begin{figure}[h!]
  \begin{minipage}[t]{0.46\textwidth}
    \centerline{\includegraphics[width=0.98\textwidth]{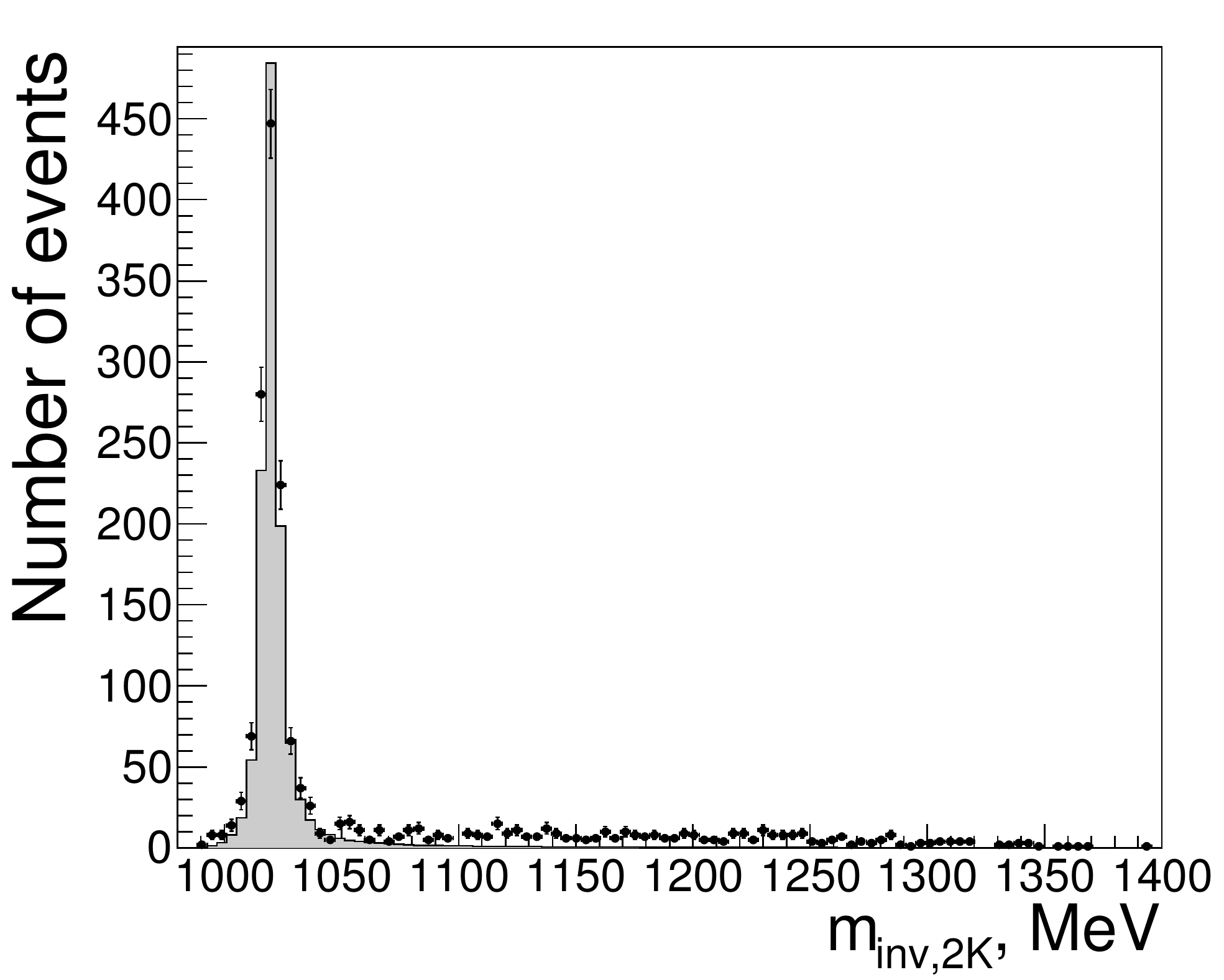}}   
    \caption{Distribution of $m_{\rm inv, 2K}$ in data (points) and simulation 
of $e^{+}e^{-}{\to}\phi\eta{\to}K^+K^-2\gamma$ (the grey histogram). 
Data at all energies are used. \label{fig:m_2K}}
  \end{minipage}\hfill\hfill  
  \begin{minipage}[t]{0.46\textwidth}
    \centerline{\includegraphics[width=0.98\textwidth]{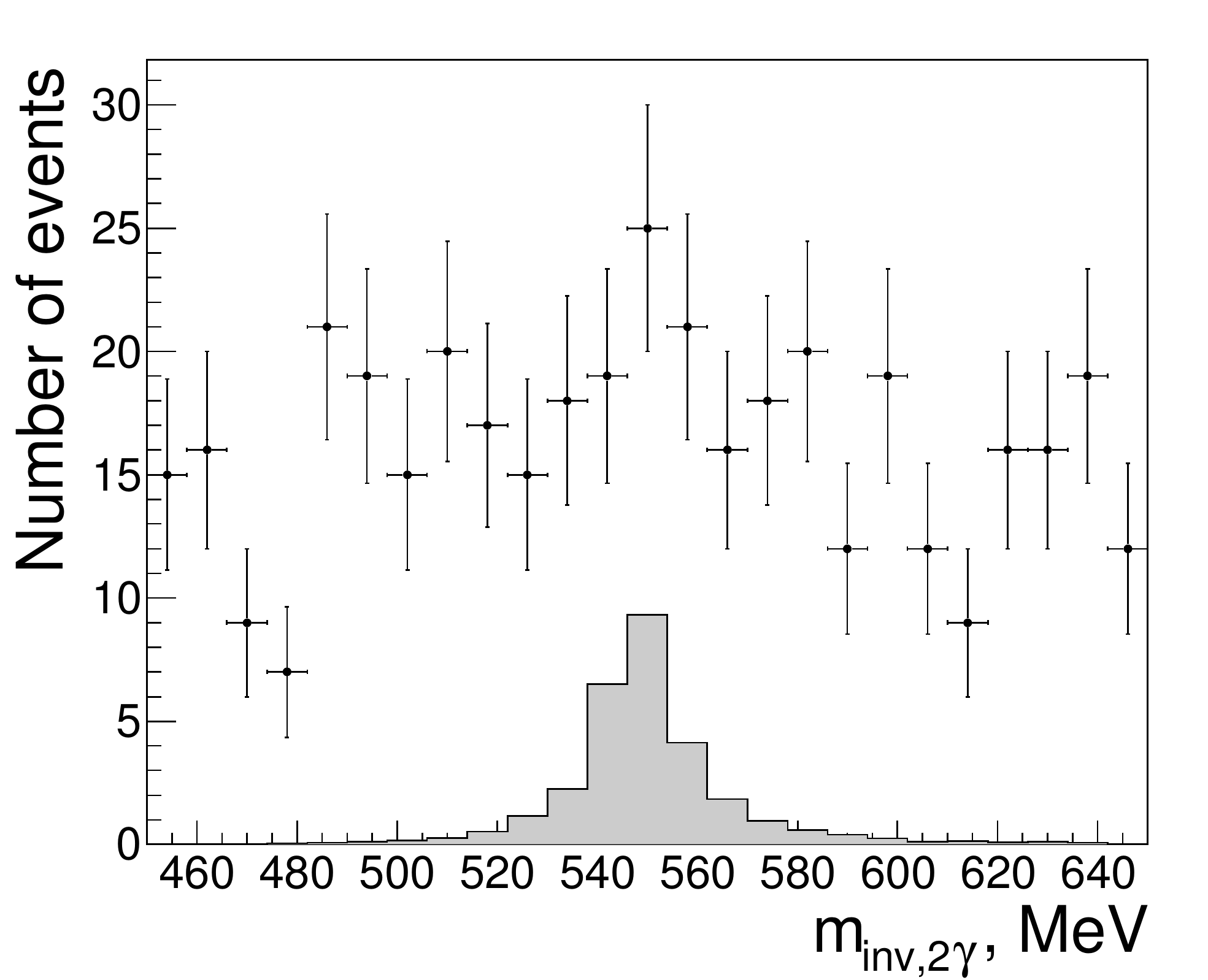}}  
    \caption{Distribution of the $m_{\rm inv, 2\gamma}$ for events with 
$m_{\rm inv, 2K}>1075 \, \rm MeV$ in data (points) and simulation of 
$e^{+}e^{-}{\to}\phi\eta{\to}K^+K^-2\gamma$ (the grey histogram). Data at all 
energies are used. \label{fig:m_2g_phi_tail}}
  \end{minipage}\hfill\hfill  
\end{figure}

The resulting distributions of $dE/dx$ vs particle momentum, 
$m_{\rm inv, 2\gamma}$ and $m_{\rm inv, 2K}$ are shown in Figs.~\ref{fig:dEdx}--\ref{fig:m_2K}. It is seen that the $\phi\eta{\to}K^{+}K^{-}\eta$ mechanism 
dominates in the process. Furthermore, events with $m_{\rm inv,\,2K}>1075\,{\rm MeV}$ show no peaking structure near $m_{\rm inv, 2\gamma}=m_{\eta}$, thus 
mainly coming from the background (the expected contribution of 
$\phi\eta{\to}K^{+}K^{-}2\gamma$ is about 30 events).

Thus, we do not observe a contribution of any intermediate states in the
process $e^+e^-{\to}K^+K^-\eta$ other than $\phi\eta$. In what follows we 
measure the cross section of the process $e^+e^-{\to}\phi\eta$ using the
recoil to an $\eta$ meson. Such an inclusive approach allows us to avoid the 
loss of statistics due to selection of specific $\eta$ decay modes, but in 
turn it increases the amount of background.


\subsection{Signal/background separation}

\begin{figure}[h!]
  \begin{minipage}[t]{0.46\textwidth}
    \centerline{\includegraphics[width=0.98\textwidth]{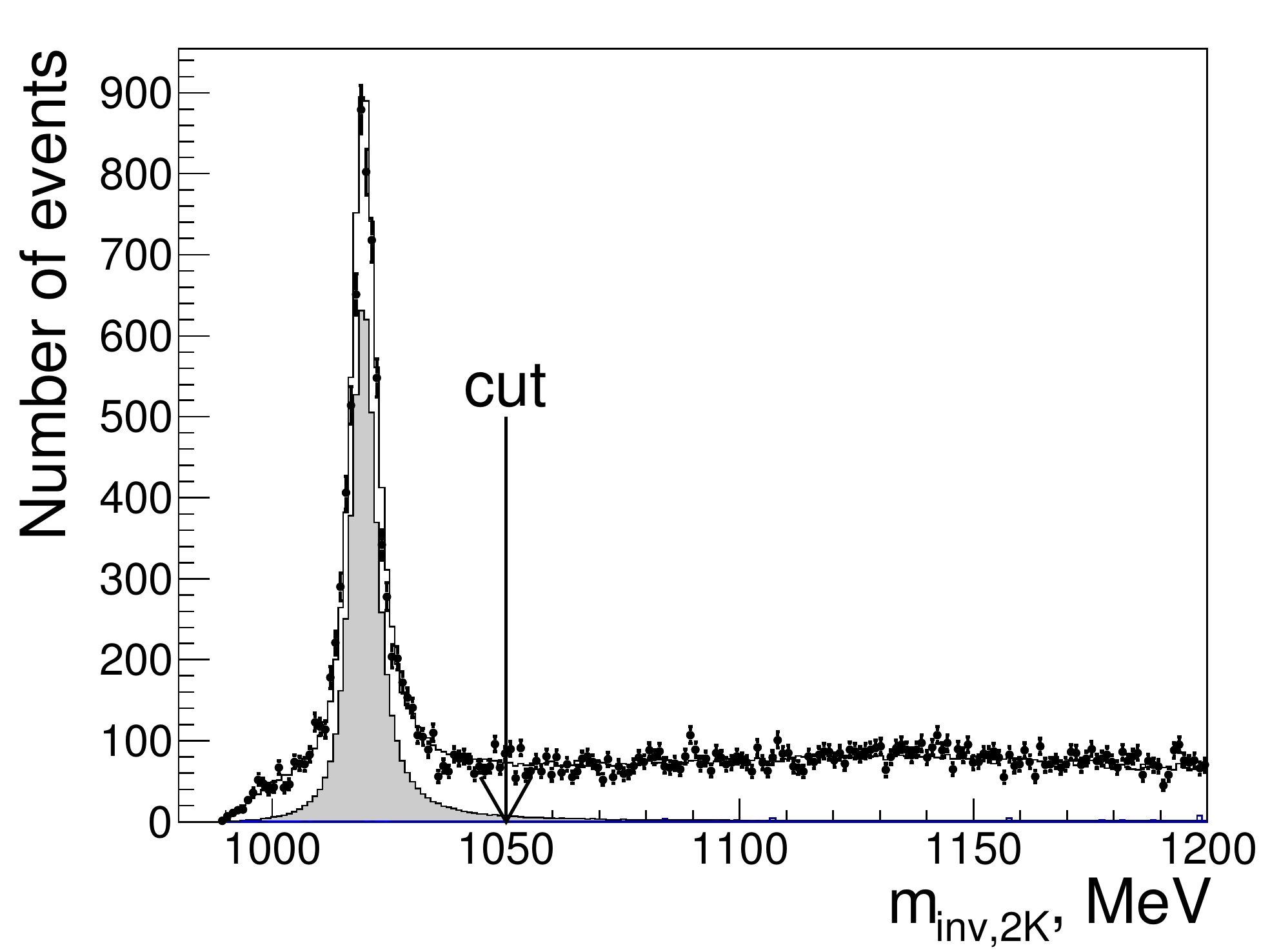}}   
    \caption{Distribution of the $m_{\rm inv, 2K}$ in data (points), 
in simulation of the signal (the grey histogram) and of the signal and background processes (the open histogram).
 Data at all energies
are used. \label{fig:m_2K_inclusive}}
  \end{minipage}\hfill\hfill  
  \begin{minipage}[t]{0.46\textwidth}
    \centerline{\includegraphics[width=0.98\textwidth]{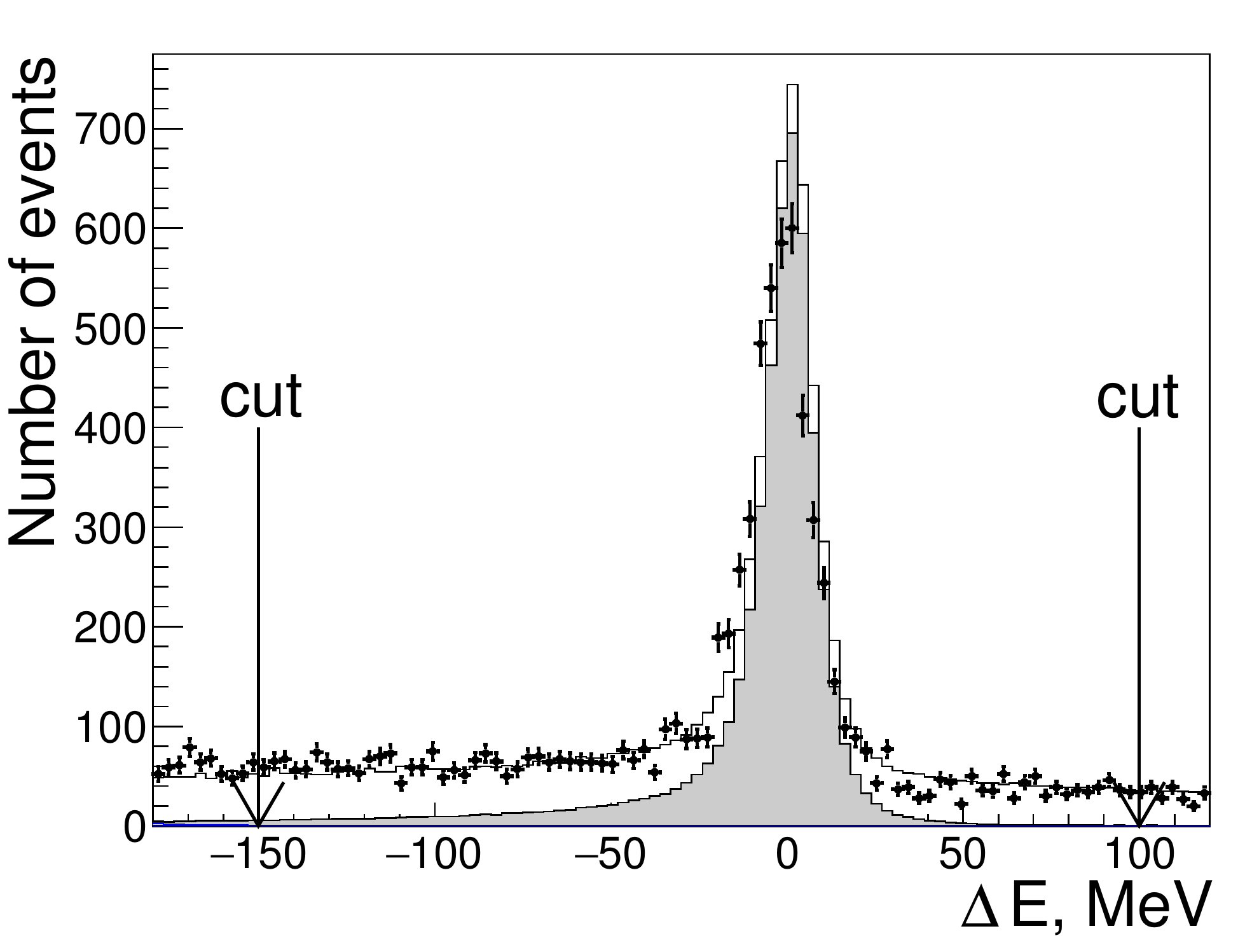}}
    \caption{Distribution of ${\Delta}E$ in data (points), 
in simulation of the signal (the grey histogram) and of the signal and background processes (the open histogram). 
Data at all energies are used. \label{fig:dE}}
  \end{minipage}\hfill\hfill  
\end{figure}

We use the requirement $L_{\rm 2K}>-0.3$ to select events with two oppositely 
charged kaons and then the requirement $m_{\rm inv,\,2K}<1050\,{\rm MeV}$ 
to select events from the $\phi$-meson region, see Fig.~\ref{fig:m_2K_inclusive}. Simulation shows that the major background final states are $K^+K^-\pi^{0}\pi^{0}$~\cite{shemyakin_kkpipi,babar_kkpipi} and $K^{+}K^{-}\pi^{+}\pi^{-}$~\cite{babar_kkpipi}.

We perform the signal/background separation using the distribution of 
the ${\Delta}E$ parameter (Fig.~\ref{fig:dE}), which is defined as

\begin{eqnarray}
  {\Delta}E=\sqrt{\vec{p}^{2}_{K^{+}}+m^{2}_{K^{+}}}+\sqrt{\vec{p}^{2}_{K^{-}}+m^{2}_{K^{-}}}+\sqrt{(\vec{p}_{K^{+}}+\vec{p}_{K^{-}})^{2}+m^{2}_{\eta}}-E_{\rm c.m.},
\end{eqnarray}

\noindent and represents the ``energy disbalance" of the event assuming the 
$\eta$ to be the recoil particle for the $K^{+}K^{-}$ pair. 

We approximate the distribution of ${\Delta}E$ in the range from --150 to 100 MeV at each $E_{\rm c.m.}$, see 
Fig.~\ref{fig:dE_one_point}. The linear function is used to describe the 
background shape. The shape of the signal is determined at each $E_{\rm c.m.}$ 
by fitting the simulated signal ${\Delta}E$ distribution by the sum of three 
Gaussians:

\begin{eqnarray}
f^{\rm MC}_{\rm sig}(x) = a_{0}\Bigl(a_{1}G(x,\mu_{1},\sigma_{1})+a_{2}G(x,\mu_{2},\sigma_{2})+(1-a_{1}-a_{2})G(x,\mu_{3},\sigma_{3})\Bigr),
\end{eqnarray}

\begin{eqnarray}
G(x,\mu,\sigma)=\frac{1}{\sqrt{2\pi}\sigma}\exp\Biggl(-\frac{(x-\mu)^{2}}{2\sigma^{2}}\Biggr). \nonumber
\end{eqnarray}

In the fit of the experimental ${\Delta}E$ distribution we fix the parameters 
$a_{1,2}$, $\mu_{i}$ and $\sigma_{i}$ characterising the signal shape at the 
values obtained from the fit of the simulation. The signal amplitude $a_{0}$, 
the possible shift ${\delta}x$ and smearing $\delta\sigma$ of the signal 
distribution are taken as floating parameters: 
\begin{eqnarray}
f^{\rm exp}_{\rm sig}(x) = a_{0}\Bigl(a_{1}G(x,\mu_{1}+{\delta}x,\sqrt{\sigma^{2}_{1}+\delta\sigma^{2}})+a_{2}G(x,\mu_{2}+{\delta}x,\sqrt{\sigma^{2}_{2}+\delta\sigma^{2}})+\\
(1-a_{1}-a_{2})G(x,\mu_{3}+{\delta}x,\sqrt{\sigma^{2}_{3}+\delta\sigma^{2}})\Bigr). \nonumber
\end{eqnarray}

In total, 3009 $\pm$ 67 of signal events were selected.


\subsection{Efficiencies\label{sec:efficiencies}}

\begin{figure}[h!]
  \begin{minipage}[t]{0.46\textwidth}
    \centerline{\includegraphics[width=0.98\textwidth]{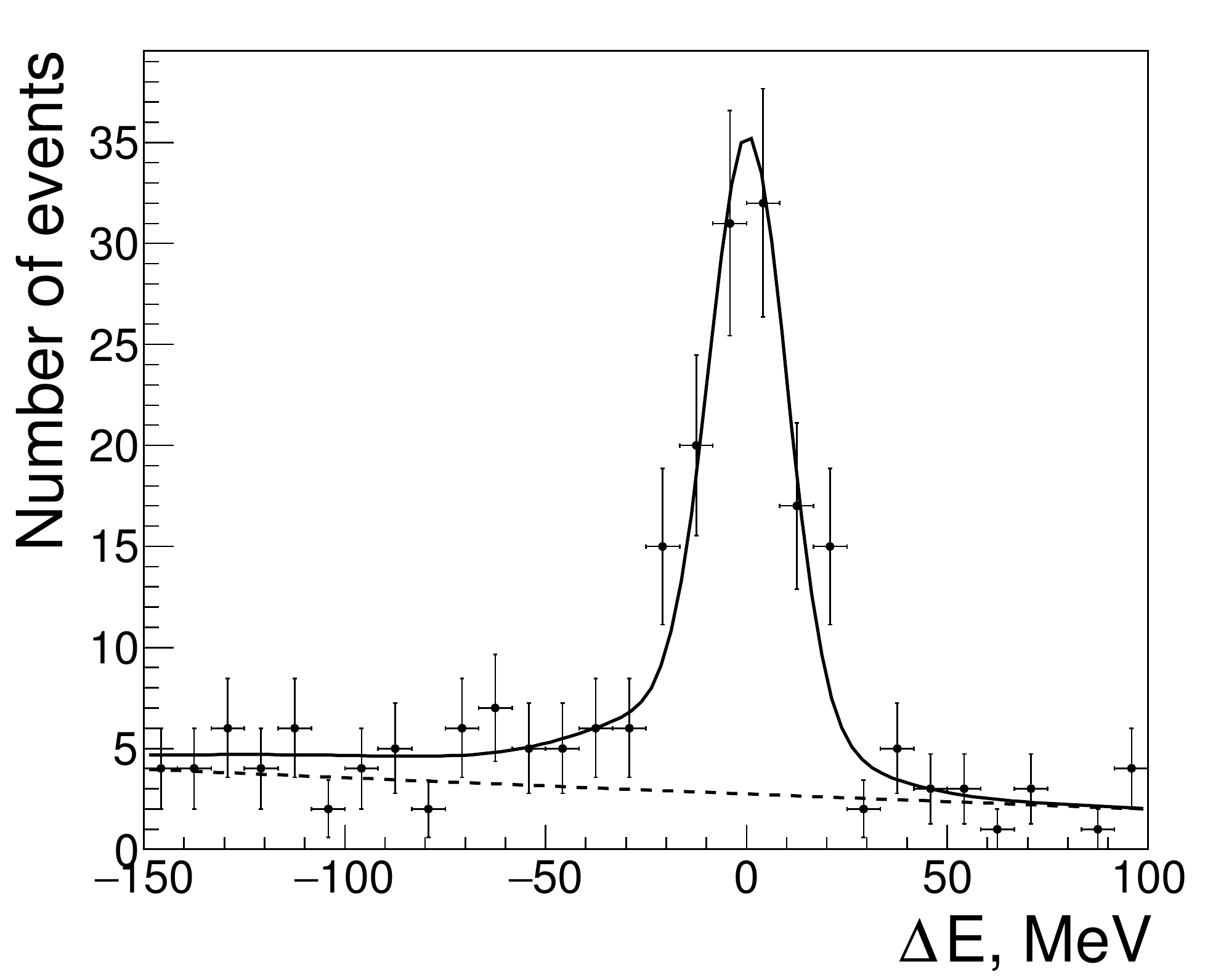}}
    \caption{Signal/background separation at $E_{\rm c.m.}=1.967\,\rm GeV$ by a
fit of the ${\Delta}E$ distribution in data (points). The solid line 
represents the fitting function, the dotted line - the part of this function 
related to the background.\label{fig:dE_one_point}}
  \end{minipage}\hfill\hfill
  \begin{minipage}[t]{0.46\textwidth}
    \centerline{\includegraphics[width=0.98\textwidth]{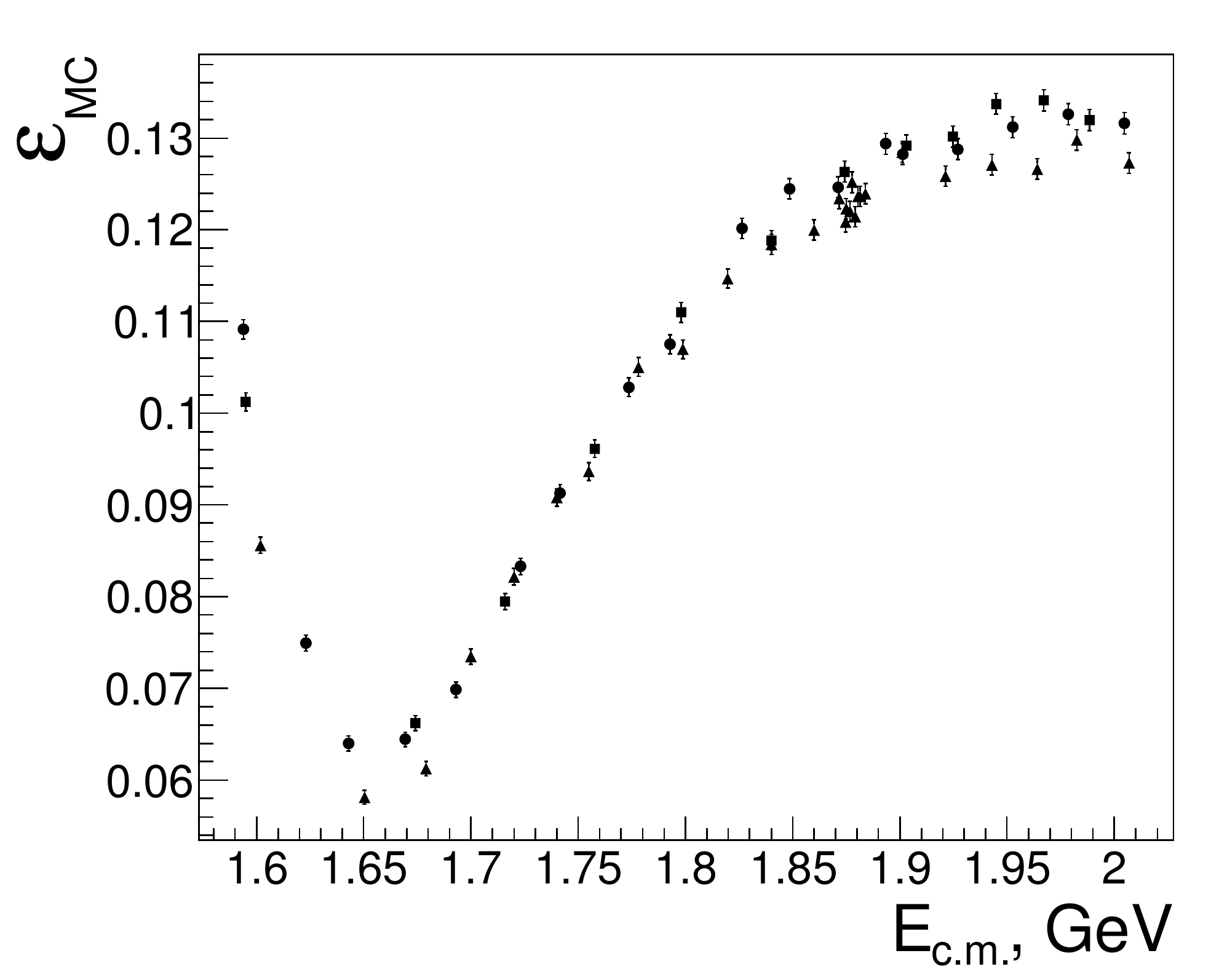}}
    \caption{Detection efficiency for events of the process 
$e^+e^-{\to}K^+K^-\eta$ as a function of $E_{\rm c.m.}$.\label{fig:eff_mc}}
  \end{minipage}\hfill\hfill
\end{figure}

Figure~\ref{fig:eff_mc} shows the detection efficiency for events of the signal
process (including emission of photon jets by the initial electron and 
positron) according to simulation ($\varepsilon_{\rm MC}$) depending on 
$E_{\rm c.m.}$, calculated as the ratio of the number of detected events
in simulation to the total number of simulated events. The nonmonotonous 
behaviour of $\varepsilon_{\rm MC}$ reflects the dependence of the geometrical 
detection efficiency of the kaon pair produced in the $\phi$-meson decay
on the $\phi$-meson velocity. The ``jumps" in $\varepsilon_{\rm MC}$ are related 
to the variation of the $dE/dx$ resolution at different energy points.

In the study of the process $e^{+}e^{-}{\to}K^{+}K^{-}\pi^{+}\pi^{-}$ with the 
CMD-3 detector~\cite{shemyakin_kkpipi} it was found that the average detection 
efficiencies for kaons in experiment, ($\varepsilon^{K}_{\rm exp}$), and 
simulation, ($\varepsilon^{K}_{\rm MC}$), agree with the accuracy of 1\% 
(the $0.85<\theta<\pi-0.85$ range was considered). Thus we estimate the 
systematic uncertainty of the kaon detection efficiency for the ``good" polar 
angle range $1.0<\theta<\pi-1.0$ as less than $1\%$. 

\begin{figure}[h!]
  \begin{minipage}[t]{0.46\textwidth}
    \centerline{\includegraphics[width=0.98\textwidth]{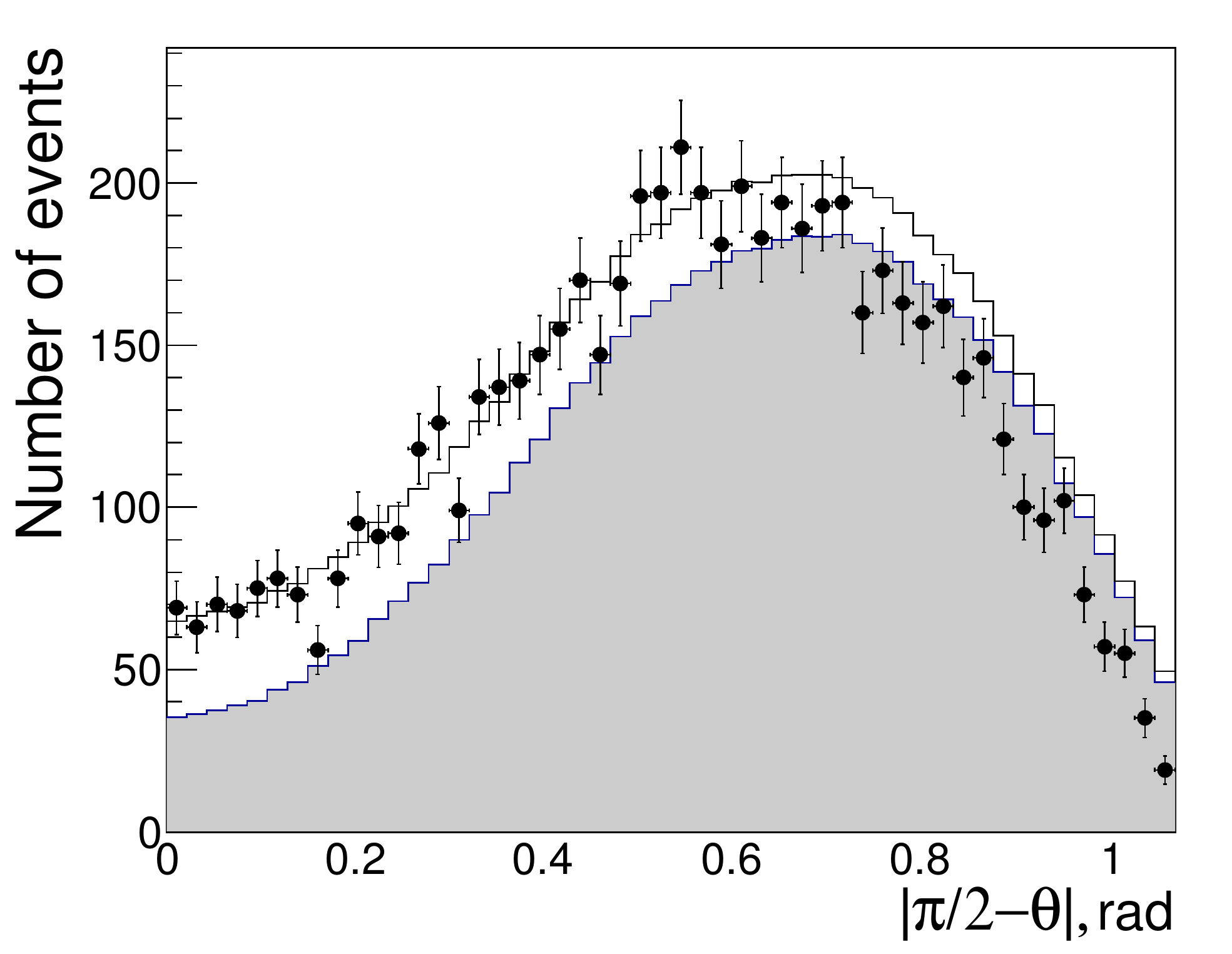}}
    \caption{Distribution of $|\pi/2-\theta|$ for the second kaon in the 
experiment and in the simulation of the signal (the grey histogram) and of the signal and background processes (the open histogram). 
The histogram for the simulation of signal and background is normalized in 
the range $|\pi/2-\theta|<0.5$ to the experimental one in the same range. \label{fig:theta}}
  \end{minipage}\hfill\hfill
  \begin{minipage}[t]{0.46\textwidth}
    \centerline{\includegraphics[width=0.98\textwidth]{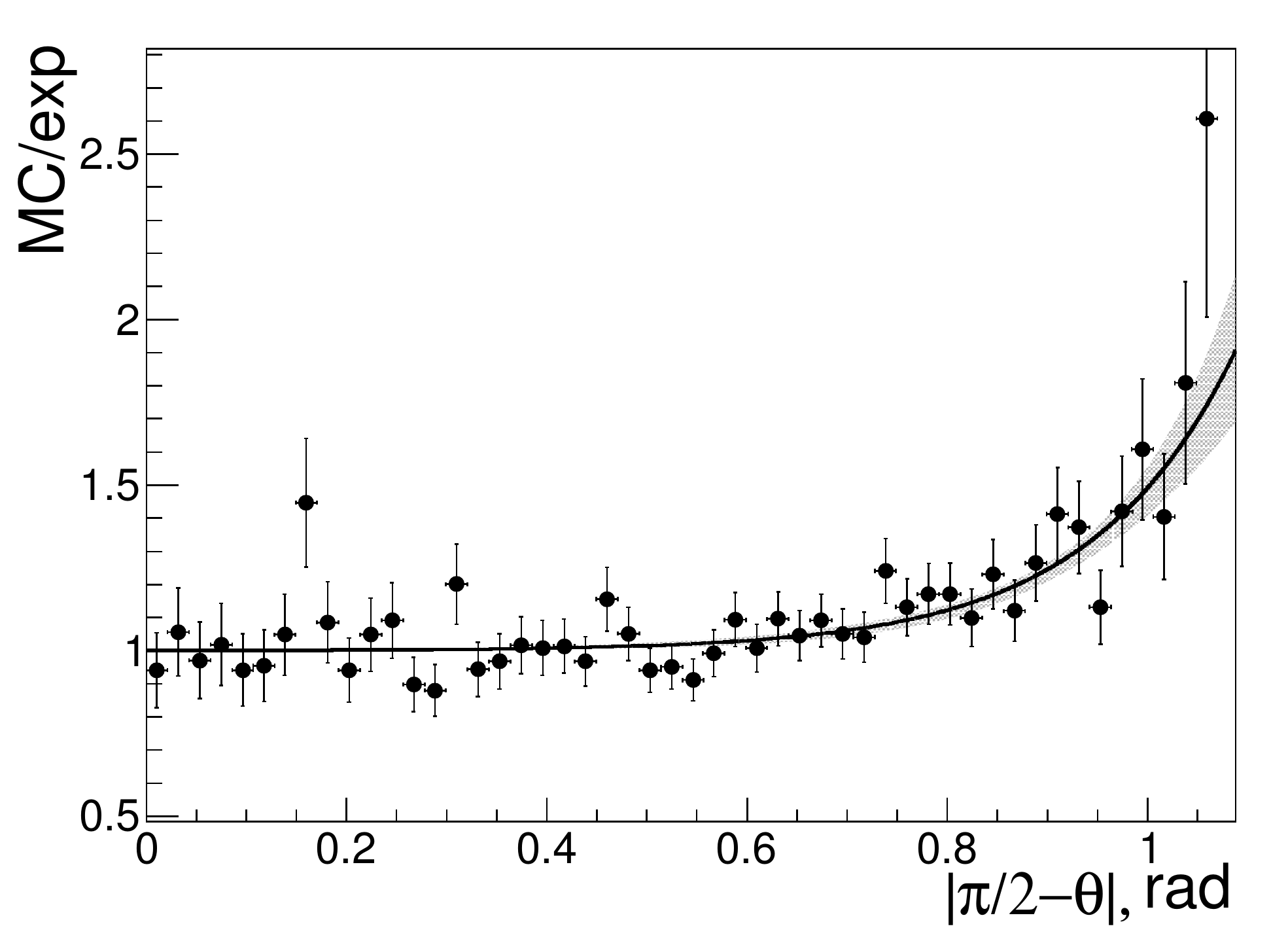}}
    \caption{Approximation of the ratio of the $|\pi/2-\theta|$ distribution 
for the second kaon in the simulation of the signal and background processes 
to that in the experiment. The shaded area shows the uncertainty of the 
fitting function. \label{fig:eff_corr_vs_theta}}
  \end{minipage}\hfill\hfill
\end{figure}

At polar angles $\theta<1.0$ and $\theta>\pi-1.0$ the kaon detection 
efficiency decreases in a different way in data and simulation, leading to 
the difference of the experimental and simulated kaon polar angle spectra. 
From that difference one can obtain the correction to the selection efficiency
for the $K^{+}K^{-}\eta$ final state. To do this we select events from the 
signal peak region $-40\, {\rm MeV} <{\Delta}E < 20\, {\rm MeV}$ with at least 
one kaon having the polar angle in the range 1.1 to $\pi-1.1$ (we assume 
$\varepsilon^{K}_{\rm exp}$ to be equal to $\varepsilon^{K}_{\rm MC}$ in this 
range). Figure~\ref{fig:theta} shows the comparison of the $|\pi/2-\theta|$ 
distributions for the second kaon in data and simulation. The approximation of 
the ratio of spectra in simulation and data  by the function 
$1+\exp(p_{0}(p_{1}-\theta))$ provides a correction for the kaon selection 
efficiency $(1+\delta^{K}_{\rm eff})(\theta)$ as a function of $\theta$, 
see Fig.~\ref{fig:eff_corr_vs_theta}. The uncertainty of this function is 
obtained by the multifold variation of the points in the histogram, shown in Fig.~\ref{fig:eff_corr_vs_theta}, 
and it's subsequent refitting. 

The correction $(1+\delta_{\rm eff})$ for the kaons selection efficiency in $K^{+}K^{-}\eta$ final state is obtained as the convolution of $1/(1+\delta^{K}_{\rm eff})(\theta)$ with the polar angle distributions of the kaons reconstructed in simulation:

\begin{eqnarray}
(1+\delta_{\rm eff})=\frac{1}{N_{\rm sim. rec.}}\sum^{i=N_{\rm sim. rec.}}_{i=1}\frac{1}{(1+\delta^{K}_{\rm eff}(\theta_{K^-})) \cdot (1+\delta^{K}_{\rm eff}(\theta_{K^+}))}.
\end{eqnarray}

\begin{figure}[h!]
  \begin{minipage}[t]{0.46\textwidth}
    \centerline{\includegraphics[width=0.98\textwidth]{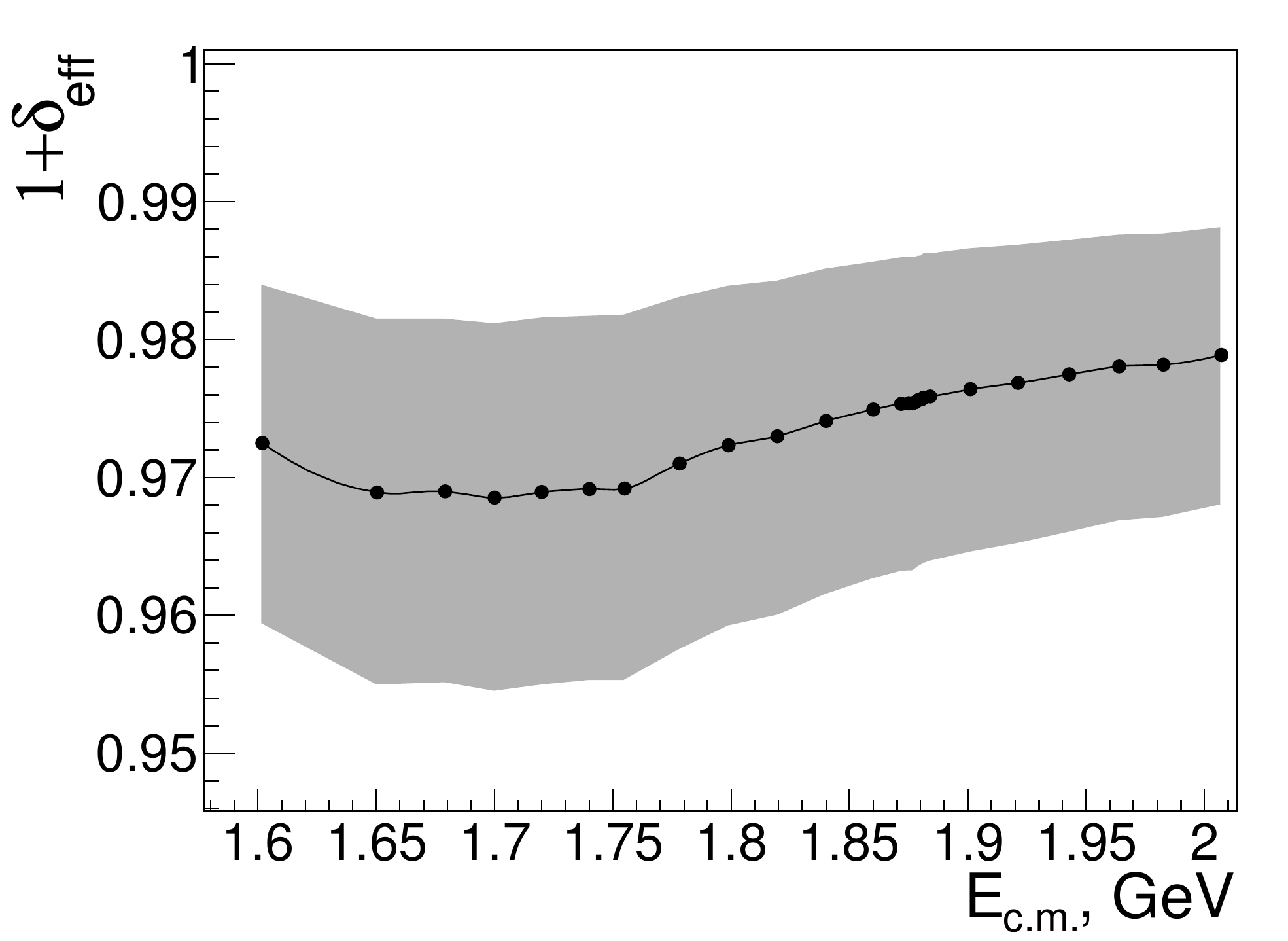}}
    \caption{Correction $(1+\delta_{\rm eff})$ to the detection efficiency of events of the
process $e^+e^-{\to}K^+K^-\eta$ as a function of $E_{\rm c.m.}$ for the runs of 
2017. The shaded area shows the uncertainty of the correction.\label{fig:eff_corr}}
  \end{minipage}\hfill\hfill
  \begin{minipage}[t]{0.46\textwidth}
    \centerline{\includegraphics[width=0.98\textwidth]{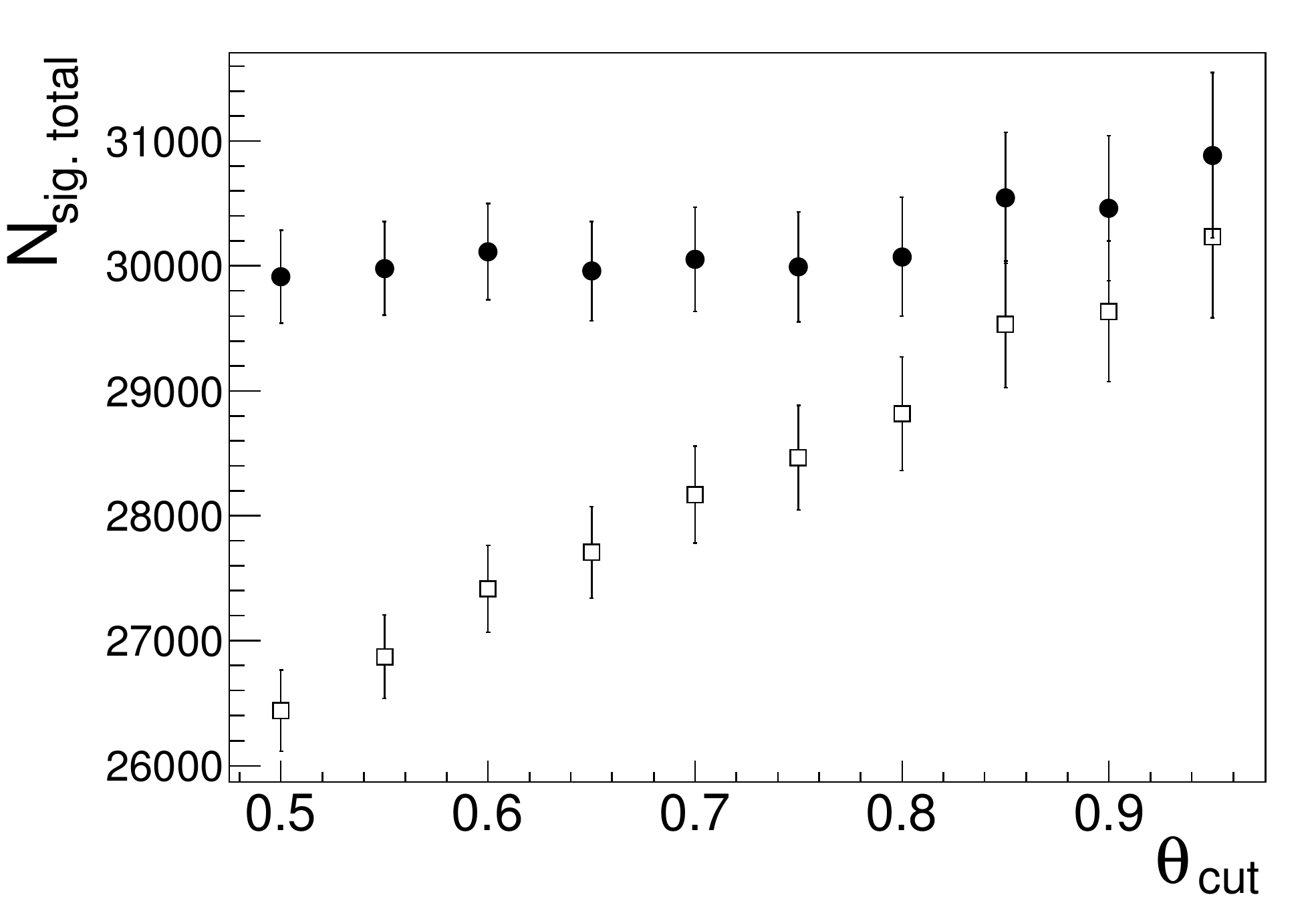}}
    \caption{Dependence of $N_{\rm sig. tot}$ on $\theta_{\rm cut}$ before 
(open bars) and after (filled circles) application of the efficiency 
correction. \label{fig:n_sig_tot}}
  \end{minipage}\hfill\hfill
\end{figure}

The values of this correction at different energies are shown in 
Fig.~\ref{fig:eff_corr}. The systematic uncertainty of these values is derived from the 
uncertainty of $(1+\delta^{K}_{\rm eff})(\theta)$ function and is estimated to be 1.5\%.
To test the validity of the obtained correction, we use the value of 
the estimated total number of signal events $N_{\rm sig. tot}$, actually produced 
at the collider during the experimental runs:

\begin{eqnarray}
  N_{\rm sig. tot}=\sum_{i=1}^{N_{\rm en.points}}\frac{N^{i}_{\rm sig. events}}{\varepsilon^{i}},
\end{eqnarray} 
where $N^{i}_{\rm sig. events}$ is the number of selected signal events at the 
$i$-th energy, $\varepsilon^{i}$ -- the corrected detection efficiency at that 
energy. Application of the efficiency correction makes $N_{\rm sig. tot}$ almost 
independent of $\theta_{\rm cut}$, as one can see from Fig.~\ref{fig:n_sig_tot}. 

Next, since the $\varepsilon_{\rm MC}$ value does not include the trigger 
efficiency $\varepsilon_{\rm trig}$, the latter should be found separately from 
the experimental data. The trigger of the CMD-3 detector consists of two 
subsystems, so-called ``neutral"\,trigger (NT) and ``charged"\,trigger (CT), 
connected into the OR scheme, and the overall trigger efficiency equals

\begin{eqnarray}
  \varepsilon_{\rm trig}=1-(1-\varepsilon_{\rm NT})(1-\varepsilon_{\rm CT}),
\end{eqnarray}
where the efficiencies of NT and CT are expressed in terms of the number 
of events in the experiment, in which only NT ($N_{\rm NT}$), only CT 
($N_{\rm CT}$) or both subsystems ($N_{\rm NT{\&}CT}$) were triggered:

\begin{eqnarray}  
  \varepsilon_{\rm NT}=\frac{N_{\rm NT\&CT}}{N_{\rm NT\&CT}+N_{\rm CT}},\,\varepsilon_{\rm CT}=\frac{N_{\rm NT\&CT}}{N_{\rm NT\&CT}+N_{\rm NT}}.
\end{eqnarray}

Figure~\ref{fig:eff_trig} shows the values of $\varepsilon_{\rm trig}$, 
$\varepsilon_{\rm NT}$ and $\varepsilon_{\rm CT}$ as the functions of $E_{\rm c.m.}$ 
for the runs of 2012. Finally, the corrected detection efficiency 
$\varepsilon$ is calculated as

\begin{eqnarray}  
  \varepsilon=\varepsilon_{\rm MC}(1+\delta_{\rm eff})\varepsilon_{\rm trig}.
\end{eqnarray}


\subsection{Cross section calculation and approximation}

The Born cross section of the process $e^{+}e^{-}{\to}\phi\eta$ at each 
$E_{\rm c.m.}$ is calculated by dividing the visible cross section 
$\sigma_{\rm vis}$ by the radiative correction $(1+\delta_{\rm rad})$:

\begin{eqnarray}
  \sigma_{\rm Born}=\frac{\sigma_{\rm vis}}{1+\delta_{\rm rad}}=\frac{N_{\rm sig. events}}{L\varepsilon(1+\delta_{\rm rad})\mathcal{B}^{\phi}_{K^+K^-}},
\end{eqnarray}
where $N_{\rm sig. events}$ is the number of selected events of the signal 
process, $L$ -- the integrated luminosity, $\varepsilon$ -- the corrected 
detection efficiency. To calculate the radiative correction at each 
$E_{\rm c.m.}$ point we use the $F(x,E_{\rm c.m.})$ structure function~\cite{kur_fad}:

\begin{eqnarray}
  1+\delta_{\rm rad}=\int\limits_{0}^{1}dx\,F(x,E_{\rm c.m.})\frac{\sigma_{\rm Born}(E_{\rm c.m.}\sqrt{1-x})}{\sigma_{\rm Born}(E_{\rm c.m.})}.\label{rad_corr}
\end{eqnarray}

We perform the calculation iteratively, using for the first iteration the 
approximation of the cross section measured by BaBar~\cite{babar_kpkmeta_2gamma}, in the $E_{\rm c.m.}$ range from 1.58 to 2.0 and from 2.3 to 3.5 GeV 
(excluding the region from 2.0 to 2.3 GeV to avoid the fitting of the $\phi(2170)$ resonance). For the cross section approximation we use the formula

\begin{eqnarray}
 \sigma_{\phi\eta}(s)=\frac{27\Gamma_{\phi}m^{2}_{\phi}}{\pi^{2}|\vec{p}_{K}(m_{\phi})|^3s}F(s){\Biggl|}\frac{a_{\rm n.r.}e^{i\Psi_{\rm n.r.}}}{s}+\sqrt{\frac{(\Gamma^{\phi^{\prime}}_{ee}\mathcal{B}^{\phi^{\prime}}_{\phi\eta})\Gamma_{\phi^{\prime}}m^{3}_{\phi^{\prime}}}{|\vec{p}_{\phi}(m_{\phi^{\prime}})|^3}}D_{\phi^{\prime}}(s){\Biggr|}^{2},
\end{eqnarray}

\begin{eqnarray}
F(s)={\int}|\vec{p}_{K^{+}}{\times}\vec{p}_{K^{-}}|^{2}\sin^{2}(\theta_{\rm normal})|D_{\phi}(p^{2}_{\phi})|^{2}d\Phi_{K^{+}K^{-}\eta}(\sqrt{s}),
\end{eqnarray}
where $D_{\phi^{\prime}}(s)=1/(s-m^{2}_{\phi^{\prime}}+i\sqrt{s}\Gamma_{\phi^{\prime}}(s))$ and $D_{\phi}(p^2_{\phi})=1/(p^{2}_{\phi}-m^{2}_{\phi}+i\sqrt{p^{2}_{\phi}}\Gamma_{\phi}(p^{2}_{\phi}))$ are the $\phi^{\prime}$ and $\phi$ propagators, 
$|\vec{p}_{\phi}(\sqrt{s})|$ is the momentum of the $\phi$ in the 
$\phi^{\prime}{\to}\phi\eta$ decay in the $\phi^{\prime}$ rest frame, 
$|\vec{p}_{K}(\sqrt{p^{2}_{\phi}})|$ is the momentum of the kaon in the 
$\phi{\to}K^{+}K^{-}$ decay in the $\phi$ rest frame, $\theta_{\rm normal}$ is 
the polar angle of the normal to the plane formed by the $\vec{p}_{K^{+}}$ and 
$\vec{p}_{K^{-}}$ vectors, $d\Phi_{K^{+}K^{-}\eta}$ is the element of three-body 
phase space. We neglect the OZI-suppressed~\cite{OZI} contribution of $\omega(1650)$, 
but consider the possible contributiuon of the resonance below reaction threshold, 
describing it via the amplitude $a_{\rm n.r.}e^{i\Psi_{\rm n.r.}}/s$.  
The factor $F(s)$ represents the ``dynamic" part of the squared matrix element averaged over the
three-body $K^{+}K^{-}\eta$ phase space.

The quantity $\Gamma_{\phi^{\prime}}(s)$ is given by the following expression 
(see~\cite{babar_kkpipi}):

\begin{eqnarray}
  \label{Gamma}
  \Gamma_{\phi^{\prime}}(s)=\Gamma_{\phi^{\prime}}\Biggl[
    \mathcal{B}^{\phi^{\prime}}_{K^{*}(892)K}\frac{\mathcal{P}_{K^{*}(892)K}(s)}{\mathcal{P}_{K^{*}(892)K}(m_{\phi^{\prime}}^{2})}+
    \mathcal{B}^{\phi^{\prime}}_{\phi\eta}\frac{\mathcal{P}_{\phi\eta}(s)}{\mathcal{P}_{\phi\eta}(m_{\phi^{\prime}}^{2})}+
    \mathcal{B}^{\phi^{\prime}}_{\phi\sigma}\frac{\mathcal{P}_{\phi\sigma}(s)}{\mathcal{P}_{\phi\sigma}(m_{\phi^{\prime}}^{2})}\Biggr],
\end{eqnarray}
where $\sigma$ designates the $f_{0}(500)$ meson, the $\mathcal{P}_{K^{*}(892)K}$ 
and $\mathcal{P}_{\phi\eta}$ functions represent the phase spaces of 
quasi-two-body final states in $\phi^{\prime}{\to}K^{*}(892)K$ and $\phi^{\prime}{\to}\phi\eta$ decays. 
According to~\cite{babar_kkpipi} we take $\mathcal{B}^{\phi^{\prime}}_{K^{*}(892)K}=0.7$, 
$\mathcal{B}^{\phi^{\prime}}_{\phi\eta}=0.2$,  
$\mathcal{B}^{\phi^{\prime}}_{\phi\sigma}=0.1$.
The $K^{*}K$ and $\phi\eta$ phase space factors have the form:

\begin{eqnarray}
  \label{1}
  \mathcal{P}_{VP}(s)=\Biggl[
    \frac{(s+m^{2}_{V}-m^{2}_{P})^{2}-4m^{2}_{V}s}{s}
    \Biggl]^{\frac{3}{2}},
\end{eqnarray}
where $V=K^{*},\phi$, $P=K,\eta$. The $\mathcal{P}_{{\phi}\sigma}$ 
function in (\ref{Gamma}) represents the phase space of the quasi-two-body final state in 
$\phi^{\prime}{\to}{\phi}\sigma$ decay and is calculated as:

\begin{eqnarray}
  \mathcal{P}_{{\phi}\sigma}(s)={\int^{\sqrt{s}-m_{\phi}}_{2m_{\pi}}}dm|BW(m,m_{\sigma},\Gamma_{\sigma})|^{2}(\mathcal{P}_{{\phi}\sigma})|_{m_{\sigma}=m},\label{formula1}
\end{eqnarray}
where 

\begin{eqnarray}
  |BW(m,m_{\sigma},\Gamma_{\sigma})|^{2}=\frac{2m_{\sigma}\Gamma_{\sigma}m}{{\pi}((m^{2}-m_{\sigma}^{2})^{2}+m_{\sigma}^{2}\Gamma_{\sigma}^{2})}
\end{eqnarray}
is the probability density for $\sigma$ to have a mass $m$, which can be 
approximately estimated as a squared module of the Breit-Wigner function 
$BW(m,m_{\sigma},\Gamma_{\sigma})$ with the $m_{\sigma}$ central value and the 
$\Gamma_{\sigma}$ width (we set $m_{\sigma}=0.475\,\rm GeV$ and 
$\Gamma_{\sigma}=0.550\,\rm GeV$~\cite{pdg19}), and

\begin{eqnarray}
  (\mathcal{P}_{{\phi}\sigma})|_{m_{\sigma}=m}=\frac{\sqrt{(s+m^{2}_{\phi}-m^{2})^{2}-4sm^{2}_{\phi}}}{s^{3/2}}{\Biggl (}1+\frac{(s+m_{\phi}^{2}-m^{2})^2}{8sm_{\phi}^{2}}{\Biggr )}
\end{eqnarray}
is the quantity proportional to the width of the 
$\phi^{\prime}{\to}{\phi}\sigma$ 
decay with the $\sigma$ mass equal to $m$. Integration in the 
formula~\ref{formula1} is performed in the range available for 
$m=m_{\rm inv,2\pi}\in(2m_{\pi};\sqrt{s}-m_{\phi})$.

Similarly to $\Gamma_{\phi^{\prime}}(s)$ the $\Gamma_{\phi}(s)$ is calculated taking into account the $K^{+}K^{-}$, $K_{S}K_{L}$ and $\pi^{+}\pi^{-}\pi^{0}$ modes of $\phi$-meson decay. 

\begin{figure}[h!]
  \begin{minipage}[t]{0.46\textwidth}
    \centerline{\includegraphics[width=0.98\textwidth]{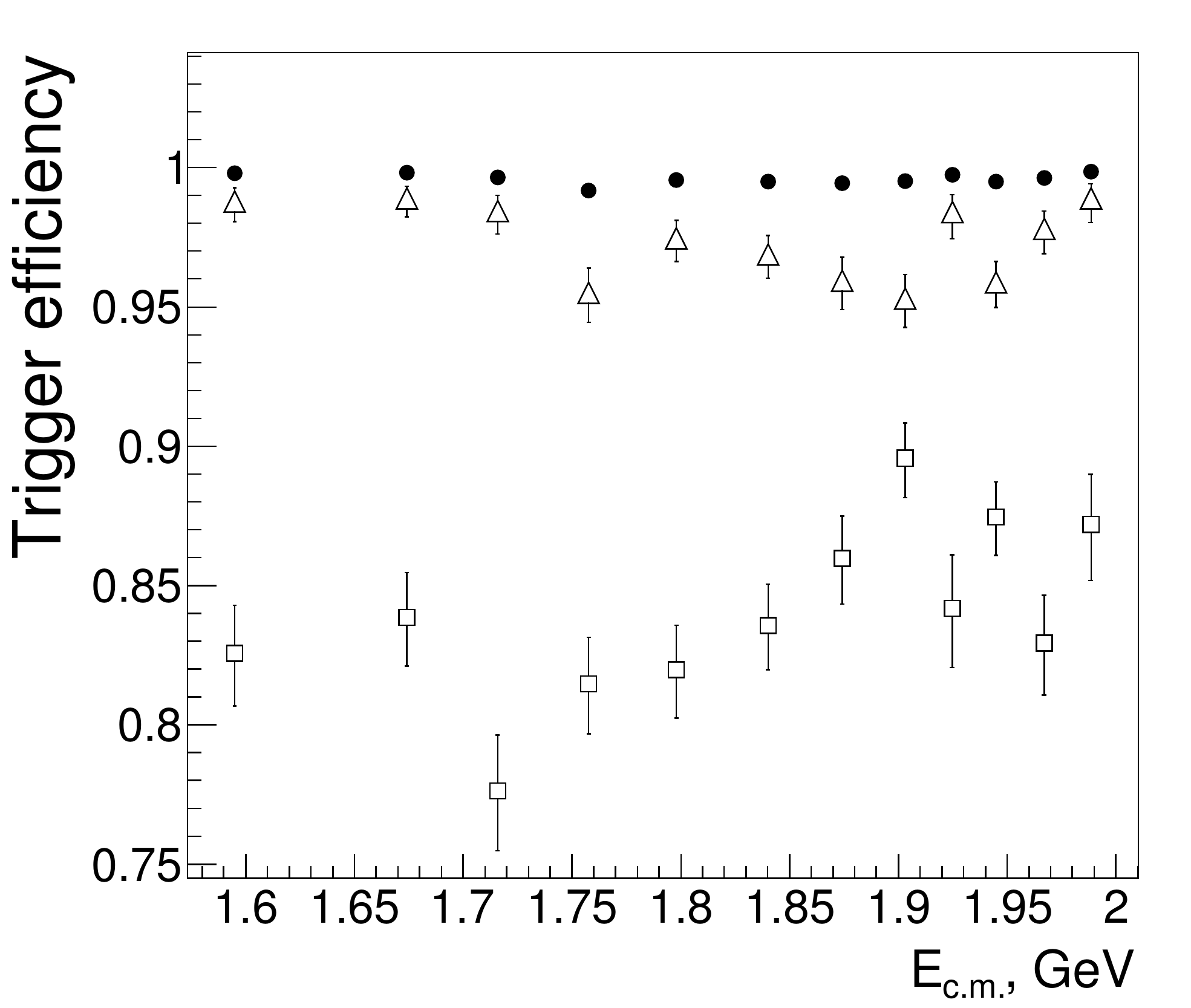}}
    \caption{The $\varepsilon_{\rm trig}$ (circles), $\varepsilon_{\rm NT}$ (open bars) and $\varepsilon_{\rm CT}$ (open triangles) values as the functions of 
$E_{\rm c.m.}$ for the runs of 2012. \label{fig:eff_trig}}
  \end{minipage}\hfill\hfill
  \begin{minipage}[t]{0.46\textwidth}
    \centerline{\includegraphics[width=0.98\textwidth]{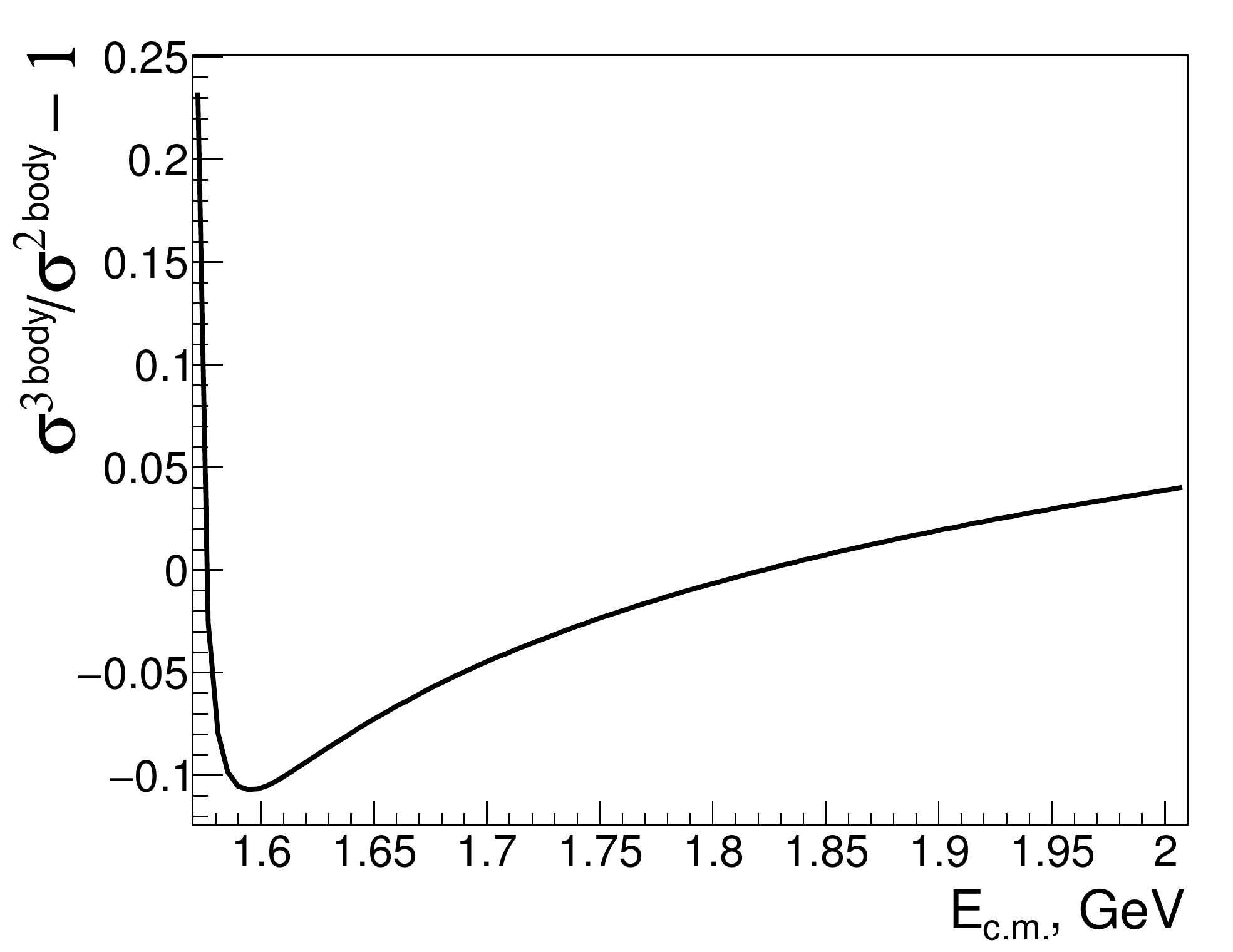}}
    \caption{Normalized difference $(\sigma^{\rm 3body}_{\phi\eta}/\sigma^{\rm 2body}_{\phi\eta}-1)$ of the two cross section parametrizations. \label{fig:cs_3b_2b_diff}}
  \end{minipage}\hfill\hfill
\end{figure}

It should be noted that in the work of BaBar~\cite{babar_kpkmeta_2gamma} for the $e^{+}e^{-}{\to}\phi\eta$ cross section fit the quasi-two-body formula

\begin{eqnarray}
  \sigma_{\phi\eta}(s)=12\pi\frac{|\vec{p}_{\phi}(\sqrt{s})|^3}{s^{3/2}}{\Biggl|}\frac{a_{\rm n.r.}e^{i\Psi_{\rm n.r.}}}{s}+\sqrt{\frac{(\Gamma^{\phi^{\prime}}_{ee}\mathcal{B}^{\phi^{\prime}}_{\phi\eta})\Gamma_{\phi^{\prime}}m^{3}_{\phi^{\prime}}}{|\vec{p}_{\phi}(m_{\phi^{\prime}})|^3}}D_{\phi^{\prime}}(s){\Biggr|}^{2}
\end{eqnarray}
was used. The normalized difference $(\sigma^{\rm 3body}_{\phi\eta}/\sigma^{\rm 2body}_{\phi\eta}-1)$ of the three-body and quasi-two-body cross section 
parametrizations is shown in Fig.~\ref{fig:cs_3b_2b_diff}. At the current 
level of a systematic uncertainty (see Section~\ref{sec:systematic}) it 
becomes important for us to use a more precise three-body formula.

\begin{figure}[h!]
  \begin{minipage}[t]{0.46\textwidth}
    \centerline{\includegraphics[width=0.98\textwidth]{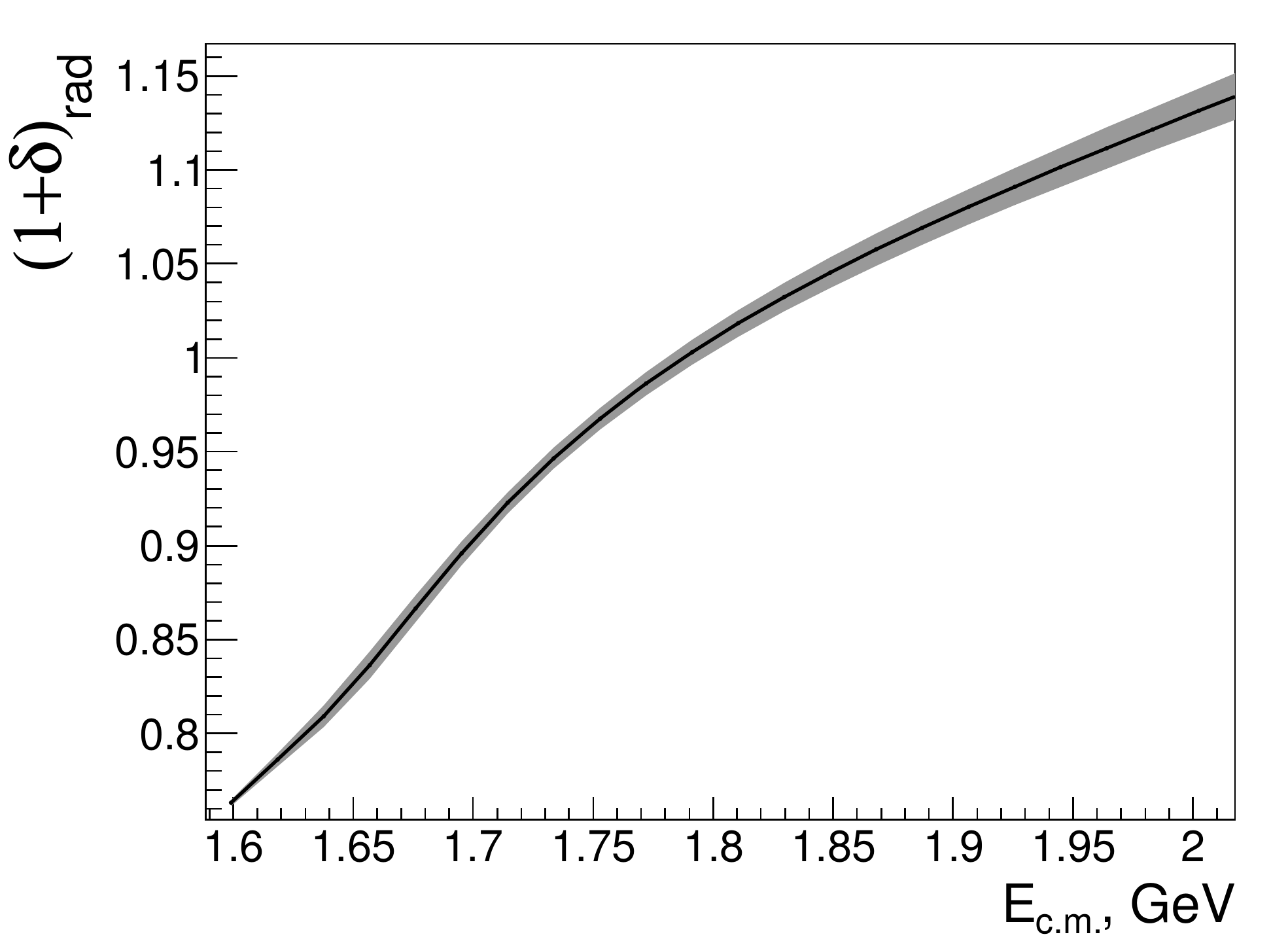}}
    \caption{The radiative correction depending on $E_{\rm c.m.}$ at the last 
iteration (the solid curve) and its uncertainty (the shaded area). \label{fig:rc}}
  \end{minipage}\hfill\hfill
  \begin{minipage}[t]{0.46\textwidth}
    \centerline{\includegraphics[width=0.98\textwidth]{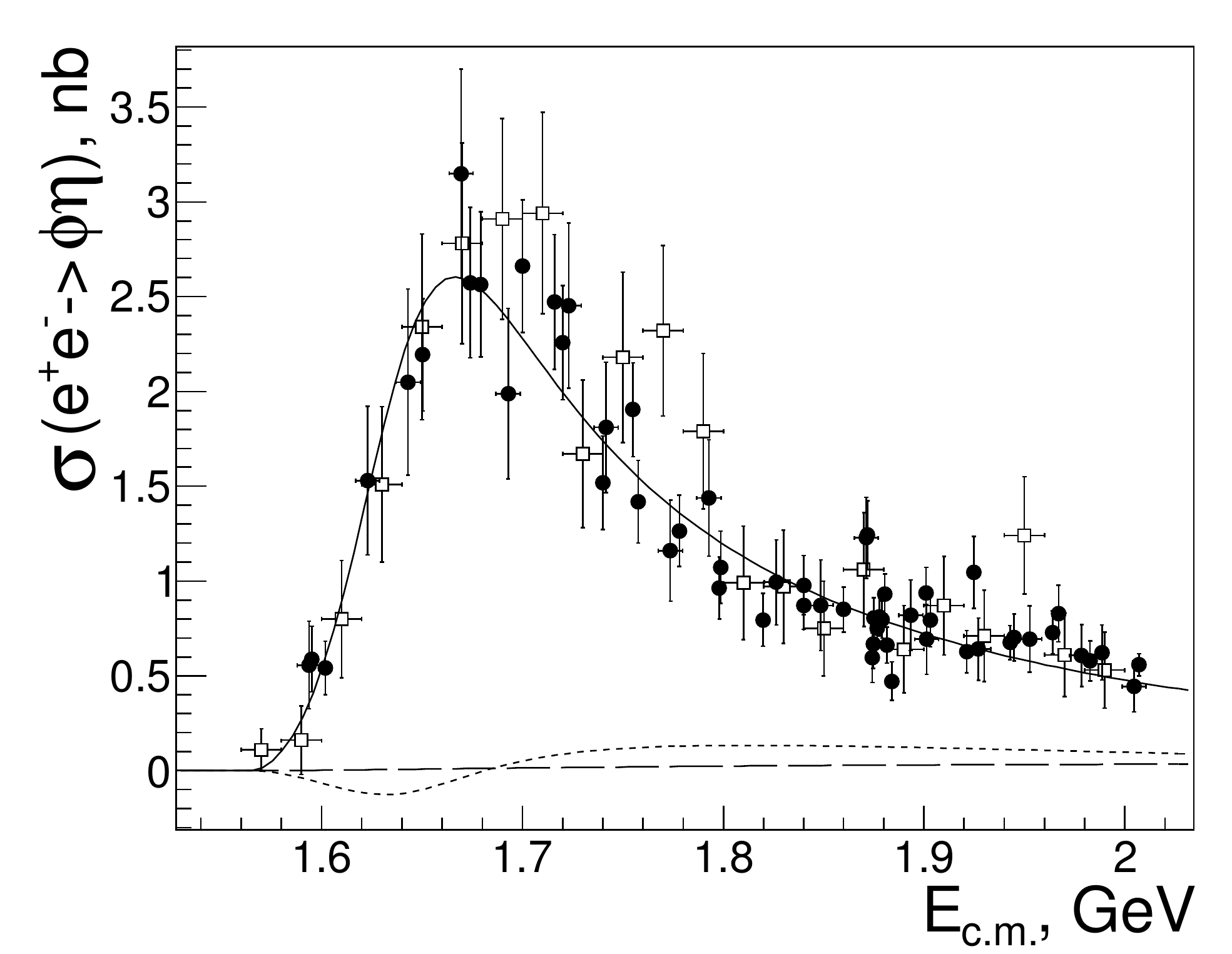}}
    \caption{BaBar (open bars), SND (open triangles) and CMD-3 (filled circles) results for the 
measurement of the $e^+e^-{\to}\phi\eta$ cross section. The overall fit of 
CMD-3 data (the solid curve), nonresonant part (the dashed curve) and 
the interference part of the fit (the dotted curve) are shown. \label{fig:cs_phieta}}
  \end{minipage}\hfill\hfill  
\end{figure}

After the first iteration we use CMD-3 data along with the BaBar data in the 
range from 2.3 to 3.5 GeV, which is necessary to fix the asymptotic behavior 
of the cross section. Four iterations are sufficient for the radiative 
corrections to converge with the accuracy of 0.5\%. Figure~\ref{fig:rc} shows 
the values of the radiative correction at the last iteration. The uncertainties
of the radiative corrections caused by the cross section 
shape are calculated by the multifold variation of the visible cross sections 
and subsequent recalculation of the radiative corrections and were found to 
be $<1.5\%$.

\begin{figure}[h!]  
  \begin{minipage}[t]{0.46\textwidth}
    \centerline{\includegraphics[width=0.98\textwidth]{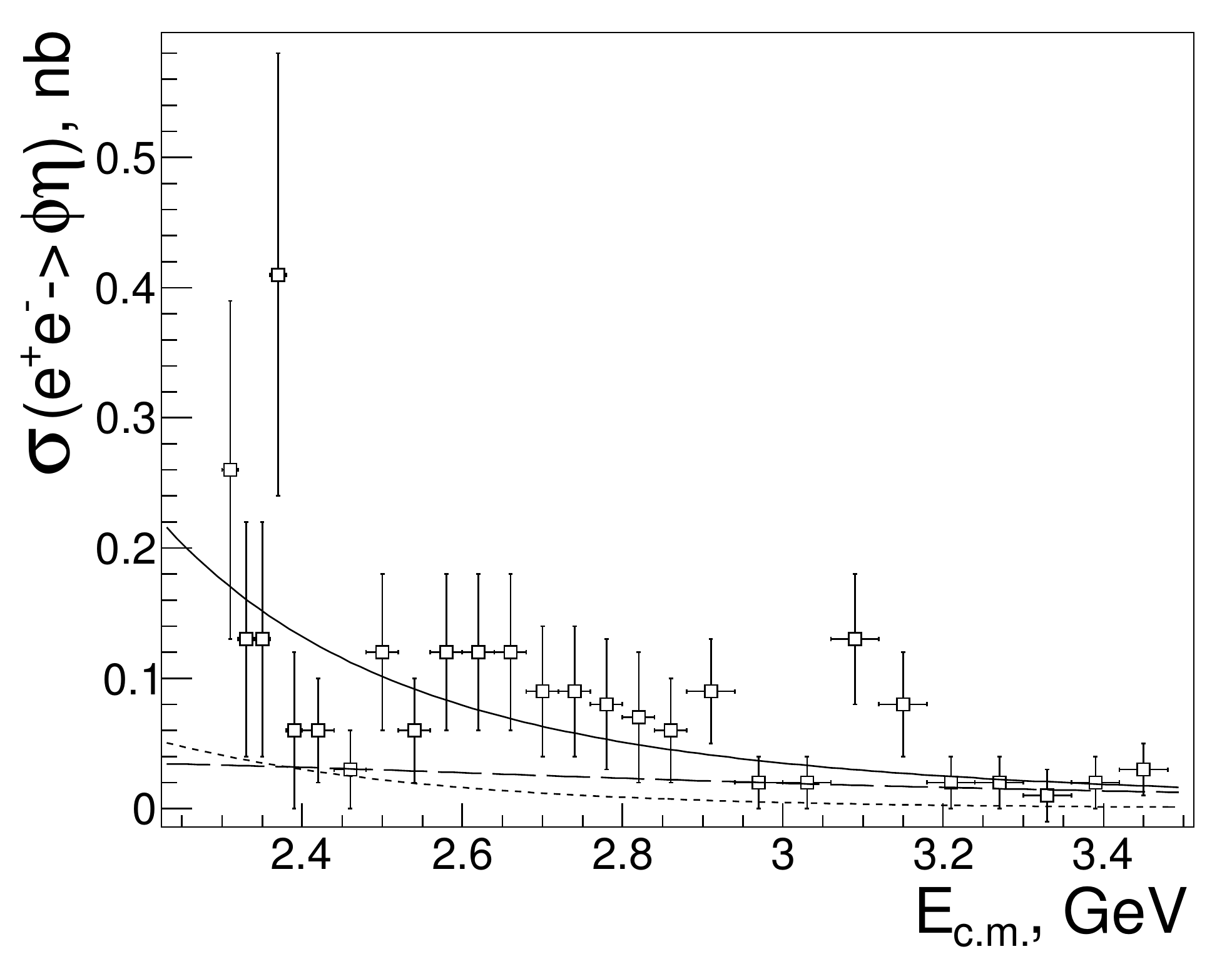}}
    \caption{Approximation of the $e^+e^-{\to}\phi\eta$ process cross section 
measured by BaBar (open bars) in the $E_{\rm c.m.}$ range from 2.3 to 3.5 GeV 
(last iteration). The overall fit (the solid curve), nonresonant part (the 
dashed curve) and the interference part of the fit (the dotted curve) are 
shown.\label{fig:cs_phieta_3_5GeV}}
  \end{minipage}\hfill\hfill
  \begin{minipage}[t]{0.46\textwidth}
   \centerline{\includegraphics[width=0.98\textwidth]{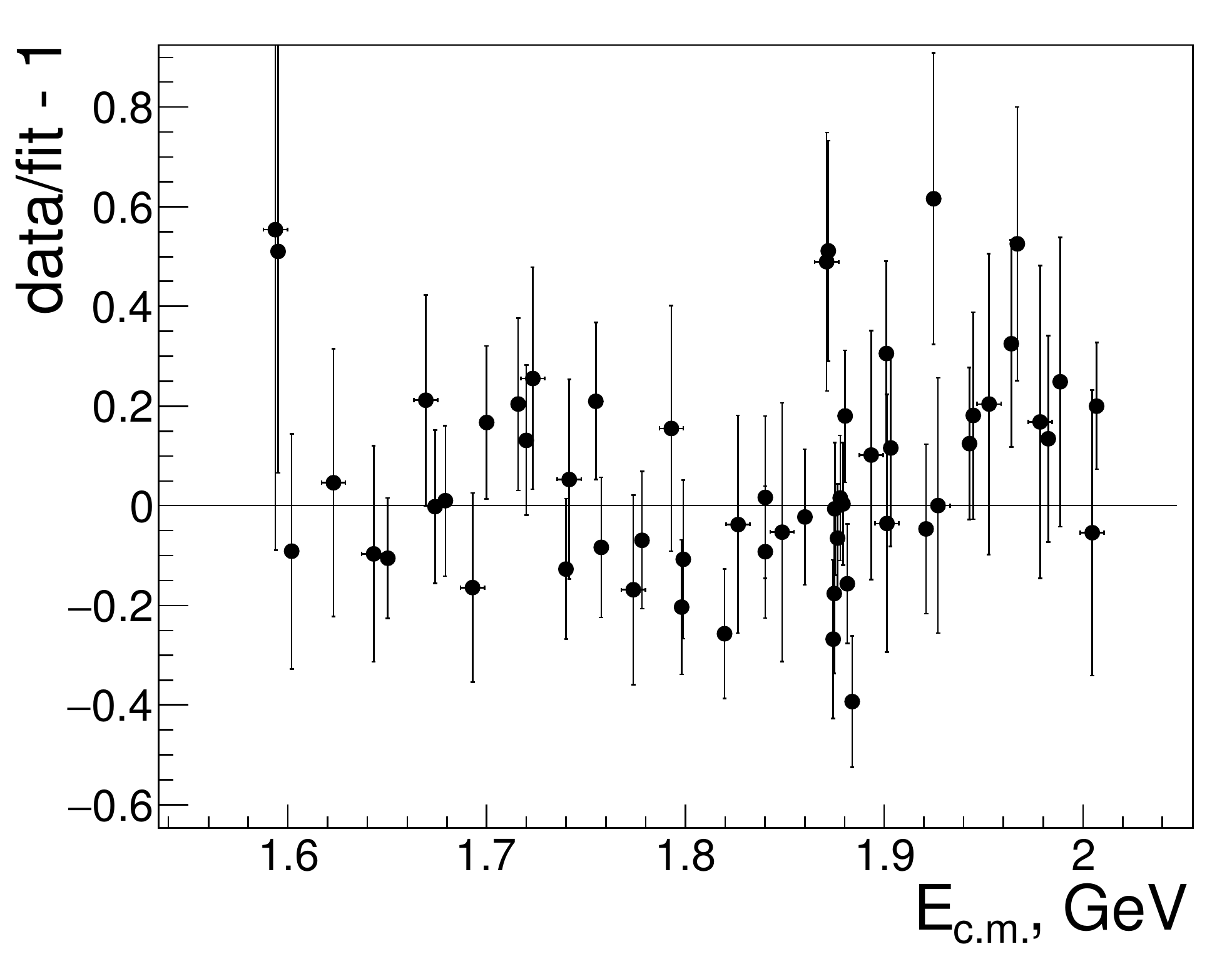}}
   \caption{Normalized difference between the $e^+e^-{\to}\phi\eta$ cross 
section measured by CMD-3 and its approximation. \label{fig:cs_diff}}
  \end{minipage}\hfill\hfill
\end{figure}

The obtained $e^{+}e^{-}{\to}\phi\eta$ Born cross section (see Tables 1--3) 
along with that of BaBar~\cite{babar_kpkmeta_2gamma} and SND~\cite{snd_kpkmeta} is shown in 
Fig.~\ref{fig:cs_phieta}. The fit of the cross section asymptotics is 
shown in Fig.~\ref{fig:cs_phieta_3_5GeV}. The obtained Born cross section 
exhibits a hint to the wavelike deviation from the fit near 
$E_{\rm c.m.}{\approx}1.9{\,}\rm GeV$, see Fig.~\ref{fig:cs_diff}. This may be 
due to the uncertainties of the branching fractions of $\phi^{\prime}$ decay modes or 
due to the decay modes, that were not taken into account 
in our cross section parameterization. Hovewer, at the current level 
of statistics we are not sensitive to these effects. 

\clearpage

\begin{table}[H]
  \begin{center}    
\caption{Center-of-mass energy $E_{\rm c.m.}$, integrated luminosity $L$, 
number of signal events $N_{\rm sig. events}$, corrected detection efficiency 
$\varepsilon$, radiative correction $(1+\delta_{\rm rad})$ and Born cross 
section of $e^{+}e^{-}{\to}\phi\eta$  for the runs of 2011. The uncertainty of $E_{\rm c.m.}$ determination is 6 MeV. Only statistical 
uncertainties are shown.}\label{tab:Table_2011}
    \begin{tabular}{cccccc}
      \hline
      $E_{\rm c.m.},\rm \,GeV$&$L$, nb$^{-1}$&$N_{\rm sig. events}$&$\varepsilon$&$1+\delta_{\rm rad}$&$\sigma$, nb\\ 
      \hline
      1.594 & 450.0 & 6.8  $\pm$ 2.7 & 0.073 $\pm$ 0.006 & 0.76 & 0.56 $\pm$ 0.23 \\
      1.623 & 518.9 & 18.4 $\pm$ 4.7 & 0.060 $\pm$ 0.002 & 0.79 & 1.53 $\pm$ 0.39 \\
      1.643 & 463.3 & 21.3 $\pm$ 5.1 & 0.056 $\pm$ 0.001 & 0.82 & 2.05 $\pm$ 0.49 \\
      1.669 & 573.2 & 41.6 $\pm$ 7.2 & 0.055 $\pm$ 0.001 & 0.86 & 3.15 $\pm$ 0.55 \\
      1.693 & 494.7 & 27.0 $\pm$ 6.1 & 0.063 $\pm$ 0.001 & 0.89 & 1.99 $\pm$ 0.45 \\
      1.723 & 531.7 & 44.2 $\pm$ 7.8 & 0.074 $\pm$ 0.001 & 0.93 & 2.45 $\pm$ 0.44 \\
      1.742 & 542.5 & 39.0 $\pm$ 7.4 & 0.085 $\pm$ 0.001 & 0.95 & 1.81 $\pm$ 0.34 \\
      1.774 & 561.6 & 29.9 $\pm$ 6.8 & 0.095 $\pm$ 0.001 & 0.98 & 1.16 $\pm$ 0.27 \\
      1.793 & 455.4 & 32.5 $\pm$ 6.9 & 0.102 $\pm$ 0.001 & 1.00 & 1.44 $\pm$ 0.31 \\
      1.826 & 514.9 & 29.1 $\pm$ 6.6 & 0.113 $\pm$ 0.001 & 1.02 & 0.99 $\pm$ 0.23 \\
      1.849 & 436.0 & 22.5 $\pm$ 6.2 & 0.117 $\pm$ 0.001 & 1.04 & 0.87 $\pm$ 0.24 \\
      1.871 & 672.8 & 50.2 $\pm$ 8.7 & 0.118 $\pm$ 0.001 & 1.05 & 1.23 $\pm$ 0.21 \\
      1.893 & 528.7 & 28.2 $\pm$ 6.4 & 0.125 $\pm$ 0.001 & 1.06 & 0.82 $\pm$ 0.19 \\
      1.901 & 506.5 & 23.4 $\pm$ 6.3 & 0.128 $\pm$ 0.001 & 1.07 & 0.69 $\pm$ 0.19 \\
      1.927 & 566.8 & 24.3 $\pm$ 6.2 & 0.126 $\pm$ 0.001 & 1.08 & 0.64 $\pm$ 0.16 \\
      1.953 & 452.0 & 21.8 $\pm$ 5.5 & 0.130 $\pm$ 0.001 & 1.09 & 0.69 $\pm$ 0.17 \\
      1.978 & 522.5 & 22.1 $\pm$ 5.9 & 0.129 $\pm$ 0.001 & 1.11 & 0.61 $\pm$ 0.16 \\
      2.005 & 481.3 & 15.3 $\pm$ 4.6 & 0.131 $\pm$ 0.001 & 1.12 & 0.44 $\pm$ 0.13 \\
    \end{tabular}
  \end{center}
\end{table}

\begin{table}[H]
  \begin{center}    
\caption{Center-of-mass energy $E_{\rm c.m.}$, integrated luminosity $L$, 
number of signal events $N_{\rm sig. events}$, corrected detection efficiency 
$\varepsilon$, radiative correction $(1+\delta_{\rm rad})$ and Born cross 
section of $e^{+}e^{-}{\to}\phi\eta$ for the runs of 2012. The uncertainty of $E_{\rm c.m.}$ determination is 2 MeV. Only statistical 
uncertainties are shown.}\label{tab:Table_2012}
    \begin{tabular}{cccccc}
      \hline
      $E_{\rm c.m.},\rm \,GeV$&$L$, nb$^{-1}$&$N_{\rm sig. events}$&$\varepsilon$&$1+\delta_{\rm rad}$&$\sigma$, nb\\ 
      \hline
      1.595 & 835.068 & 14.4 $\pm$ 4.1 & 0.076 $\pm$ 0.002 & 0.76 & 0.62 $\pm$ 0.18 \\
      1.674 & 896.135 & 57.1 $\pm$ 8.8 & 0.059 $\pm$ 0.001 & 0.87 & 2.57 $\pm$ 0.40 \\
      1.716 & 815.996 & 66.5 $\pm$ 9.5 & 0.073 $\pm$ 0.001 & 0.92 & 2.47 $\pm$ 0.35 \\
      1.758 & 972.844 & 60.2 $\pm$ 9.2 & 0.092 $\pm$ 0.001 & 0.97 & 1.42 $\pm$ 0.22 \\
      1.798 & 999.604 & 48.8 $\pm$ 8.3 & 0.103 $\pm$ 0.001 & 1.00 & 0.96 $\pm$ 0.16 \\
      1.840 & 967.496 & 55.3 $\pm$ 8.9 & 0.116 $\pm$ 0.001 & 1.03 & 0.98 $\pm$ 0.16 \\
      1.874 & 857.024 & 32.7 $\pm$ 7.1 & 0.124 $\pm$ 0.001 & 1.05 & 0.60 $\pm$ 0.13 \\
      1.903 & 901.701 & 47.6 $\pm$ 8.4 & 0.127 $\pm$ 0.001 & 1.07 & 0.79 $\pm$ 0.14 \\
      1.925 & 567.388 & 41.2 $\pm$ 7.4 & 0.131 $\pm$ 0.001 & 1.08 & 1.05 $\pm$ 0.19 \\
      1.945 & 995.035 & 47.3 $\pm$ 8.3 & 0.127 $\pm$ 0.001 & 1.09 & 0.70 $\pm$ 0.12 \\
      1.967 & 693.468 & 41.0 $\pm$ 7.4 & 0.132 $\pm$ 0.001 & 1.10 & 0.83 $\pm$ 0.15 \\
      1.988 & 601.598 & 26.9 $\pm$ 6.2 & 0.132 $\pm$ 0.001 & 1.11 & 0.62 $\pm$ 0.14 \\
    \end{tabular}
  \end{center}
\end{table}

\begin{table}[H]
  \begin{center}
\caption{Center-of-mass energy $E_{\rm c.m.}$, integrated luminosity $L$, 
number of signal events $N_{\rm sig. events}$, corrected detection efficiency 
$\varepsilon$, radiative correction $(1+\delta_{\rm rad})$ and Born cross 
section of $e^{+}e^{-}{\to}\phi\eta$ for the runs of 2017. The uncertainty of $E_{\rm c.m.}$ determination is 50 keV. Only statistical 
uncertainties are shown.}\label{tab:Table_2017}
    \begin{tabular}{cccccc}
      \hline
      $E_{\rm c.m.},\rm \,GeV$&$L$, nb$^{-1}$&$N_{\rm sig. events}$&$\varepsilon$&$1+\delta_{\rm rad}$&$\sigma$, nb\\ 
      \hline
      1.602 & 1275.5 & 18.3   $\pm$ 4.7 & 0.071   $\pm$ 0.001 & 0.77 & 0.54 $\pm$ 0.14 \\
      1.650 & 1428.8 & 65.6   $\pm$ 8.8 & 0.052   $\pm$ 0.001 & 0.83 & 2.19 $\pm$ 0.30 \\
      1.679 & 1009.5 & 60.1   $\pm$ 8.9 & 0.054   $\pm$ 0.001 & 0.87 & 2.56 $\pm$ 0.38 \\
      1.700 & 947.0  & 73.0   $\pm$ 9.5 & 0.066   $\pm$ 0.001 & 0.90 & 2.66 $\pm$ 0.35 \\
      1.720 & 923.6  & 71.7   $\pm$ 9.5 & 0.076   $\pm$ 0.001 & 0.93 & 2.26 $\pm$ 0.30 \\
      1.740 & 947.4 & 55.8    $\pm$ 9.0 & 0.083   $\pm$ 0.001 & 0.95 & 1.52 $\pm$ 0.25 \\
      1.755 & 1048.4 & 82.9   $\pm$ 10.8 & 0.088  $\pm$ 0.001 & 0.97 & 1.90 $\pm$ 0.25 \\
      1.778 & 1139.7 & 67.2   $\pm$ 9.9 & 0.097   $\pm$ 0.001 & 0.99 & 1.26 $\pm$ 0.19 \\
      1.799 & 880.9  & 48.0   $\pm$ 8.6 & 0.103   $\pm$ 0.001 & 1.00 & 1.07 $\pm$ 0.19 \\
      1.820 & 1161.7 & 50.2   $\pm$ 8.7 & 0.109   $\pm$ 0.001 & 1.02 & 0.79 $\pm$ 0.14 \\
      1.840 & 1378.3 & 68.6   $\pm$ 10.0 & 0.113  $\pm$ 0.001 & 1.03 & 0.87 $\pm$ 0.13 \\
      1.860 & 1550.5 & 78.9   $\pm$ 11.0 & 0.117  $\pm$ 0.001 & 1.05 & 0.85 $\pm$ 0.12 \\
      1.872 & 1055.5 & 80.5   $\pm$ 11.7 & 0.119  $\pm$ 0.001 & 1.05 & 1.24 $\pm$ 0.18 \\
      1.875 & 1088.3 & 44.6   $\pm$ 8.6 & 0.119   $\pm$ 0.001 & 1.05 & 0.67 $\pm$ 0.13 \\
      1.875 & 1900.0 & 94.0   $\pm$ 12.6 & 0.119  $\pm$ 0.001 & 1.05 & 0.80 $\pm$ 0.11 \\
      1.877 & 2538.6 & 117.1  $\pm$ 13.5 & 0.119  $\pm$ 0.001 & 1.05 & 0.75 $\pm$ 0.09 \\
      1.878 & 2063.5 & 103.8  $\pm$ 12.7 & 0.120  $\pm$ 0.001 & 1.06 & 0.81 $\pm$ 0.10 \\
      1.879 & 2024.8 & 99.3   $\pm$ 12.2 & 0.119  $\pm$ 0.001 & 1.06 & 0.80 $\pm$ 0.10 \\
      1.880 & 1907.2 & 110.7  $\pm$ 12.4 & 0.121  $\pm$ 0.001 & 1.06 & 0.93 $\pm$ 0.10 \\
      1.881 & 1874.3 & 78.7   $\pm$ 11.2 & 0.122  $\pm$ 0.001 & 1.06 & 0.66 $\pm$ 0.09 \\
      1.884 & 1341.7 & 39.5   $\pm$ 8.6 & 0.121   $\pm$ 0.001 & 1.06 & 0.47 $\pm$ 0.10 \\
      1.901 & 1179.9 & 71.5   $\pm$ 10.1 & 0.124  $\pm$ 0.001 & 1.07 & 0.94 $\pm$ 0.13 \\
      1.921 & 1354.4 & 55.6   $\pm$ 9.9 & 0.124   $\pm$ 0.001 & 1.08 & 0.63 $\pm$ 0.11 \\
      1.943 & 1787.7 & 78.8   $\pm$ 10.7 & 0.123  $\pm$ 0.001 & 1.09 & 0.68 $\pm$ 0.09 \\
      1.964 & 1326.1 & 65.0   $\pm$ 10.2 & 0.125  $\pm$ 0.001 & 1.10 & 0.73 $\pm$ 0.11 \\
      1.983 & 1254.5 & 49.5   $\pm$ 9.0 & 0.126   $\pm$ 0.001 & 1.11 & 0.58 $\pm$ 0.11 \\
      2.007 & 3809.4 & 143.5  $\pm$ 15.1 & 0.123  $\pm$ 0.001 & 1.12 & 0.56 $\pm$ 0.06 \\
    \end{tabular}
  \end{center}
\end{table}

The $\phi^{\prime}$ parameters, obtained from the approximation of the CMD-3 cross section are shown in Table~\ref{fit_results}.
Along with the cross section parametrization using $\Gamma^{\phi^{\prime}}_{ee}\mathcal{B}^{\phi^{\prime}}_{\phi\eta}$ we also tried the parametrization through $\mathcal{B}^{\phi^{\prime}}_{e^{+}e^{-}}\mathcal{B}^{\phi^{\prime}}_{\phi\eta}$. The results for all other fit parameters but $\Gamma^{\phi^{\prime}}_{ee}\mathcal{B}^{\phi^{\prime}}_{\phi\eta}$ and $\mathcal{B}^{\phi^{\prime}}_{e^{+}e^{-}}\mathcal{B}^{\phi^{\prime}}_{\phi\eta}$ are the same in both cases. Our results for $\phi^{\prime}$ parameters are compatible with those of BaBar~\cite{babar_kpkmeta_2gamma} 
and other previous measurements, but have better statistical precision. 
The estimation of the systematic uncertainties of $\phi^{\prime}$ parameters is described in Section~\ref{sec:systematic}.

\begin{table}[H]
  \begin{center}    
    \caption{Results of the $e^{+}e^{-}{\to}\phi\eta$ cross section approximation. \label{fit_results}}
    \begin{tabular}{ccc}      
      \hline
      Parametrization using & $\Gamma^{\phi^{\prime}}_{ee}\mathcal{B}^{\phi^{\prime}}_{\phi\eta}$ & $\mathcal{B}^{\phi^{\prime}}_{e^{+}e^{-}}\mathcal{B}^{\phi^{\prime}}_{\phi\eta}$ \\
      \hline
      Parameter & \multicolumn{2}{c}{Value} \\ 
      \hline
      $\chi^{2}/{\rm n.d.f}$ & \multicolumn{2}{c}{$93.8/79{\approx}1.19$} \\
      $\Gamma^{\phi^{\prime}}_{ee}\mathcal{B}^{\phi^{\prime}}_{\phi\eta},\rm eV$ & $94{\pm}13_{\rm stat}{\pm}15_{\rm syst}$ & -- \\
      $\mathcal{B}^{\phi^{\prime}}_{e^{+}e^{-}}\mathcal{B}^{\phi^{\prime}}_{\phi\eta}$ & -- & $0.53{\pm}0.06_{\rm stat}{\pm}0.09_{\rm syst}$ \\
      $m_{\phi^{\prime}},\rm MeV$ & \multicolumn{2}{c}{$1667{\pm}5_{\rm stat}{\pm}11_{\rm syst}$} \\
      $\Gamma_{\phi^{\prime}},\rm MeV$ & \multicolumn{2}{c}{$176{\pm}23_{\rm stat}{\pm}38_{\rm syst}$} \\
      $a_{\rm n.r.},\rm MeV$ & \multicolumn{2}{c}{$1.1{\pm}0.6_{\rm stat}$} \\
      $\Psi_{\rm n.r.}$ & \multicolumn{2}{c}{$0.14{\pm}0.67_{\rm stat}$} \\
      \hline 
    \end{tabular}
  \end{center}
\end{table}


\subsection{Systematic uncertainties \label{sec:systematic}}

We estimate a systematic uncertainty related to some selection criterion as a 
relative variation of the $N_{\rm sig. tot}$ (see Section~\ref{sec:efficiencies}) 
with the variation (or swithcing on/off) of this criterion. 
The limits for the variation of the cuts are chosen as wide as possible with two requirements: 
1) restriction does not seriously cut the signal; 
2) the background shape is reasonably described by the contribution of $K^{+}K^{-}\pi^{0}\pi^{0}$ and $K^{+}K^{-}\pi^{+}\pi^{-}$ final states.
The following sources of systematic uncertainties were considered:

\begin{itemize}
\item{
The requirements on $\rho_{\rm PCA}$, $z_{\rm PCA}$, $p_{\perp}$ and 
$dE/dx<(dE/dx)_{\rm protons}$ for positively charged particles applied in the 
``good" track selection procedure, give the uncertainties of 1.0, 0.5, 0.3 and 0.4$\%$, respectively. The values are estimated by swithcing on/off these 
requirements.
}
\item{
The cut on $L_{\rm 2K}$ used for the kaon selection was varied from -0.6 to -0.1. The corresponding uncertainty was 0.8$\%$.
}
\item{
The cut on $m_{\rm inv, 2K}$, used for the $\phi$-meson region selection, was varied from 1050 to 1100 MeV. The corresponding uncertainty was 0.7$\%$.
}
\item{
The lower limit of the ${\Delta}E$ distribution fit was varied from $-180$ to $-100$ MeV. The corresponding uncertainty was 1$\%$.
}
\item{
The upper limit of the ${\Delta}E$ distribution fit was varied from $50$ to $150$ MeV. The corresponding uncertainty was 1$\%$.
}     
\item{
The signal peak position can be fixed from simulation ($\delta x \equiv 0$) 
or released in the fit of the experimental $\Delta E$ distribution, 
the corresponding uncertainty is 2$\%$.
}
\item{
The signal width can be fixed from the simulation ($\delta \sigma \equiv 0$) 
or released, the corresponding uncertainty is 2.5$\%$.
}
\item{
The background shape in the fit of the experimental $\Delta E$ distribution 
can be taken as linear with floating parameters, or it can be fixed from the 
fit of the simulated background distribution. The corresponding uncertainty 
is 2.3$\%$.
}
\item{
The uncertainty of the single kaon detection efficiency is estimated to be 
1$\%$, for the pair of kaons -- 1.5$\%$. The uncertainty of the correction to 
the $K^{+}K^{-}\eta$ selection efficiency related to the angular dependence 
of the kaon detection efficiency (see Section~\ref{sec:efficiencies}), 
was estimated to be 1.5$\%$.  
}
\item{
The systematic uncertainty of the luminosity measurement is 1$\%$~\cite{lum}.
}
\item{
The uncertainty of the $\mathcal{B}^{\phi}_{K^+K^-}$ is about 1$\%$.
}
\end{itemize}

Table~\ref{tab:syst} shows a summary of the analyzed systematic uncertainties 
of the cross section measurement. The overall systematic uncertainty is 
obtained by a quadratic summation of the individual uncertainties and is 
estimated to be 5.1$\%$.

The following contributions to the systematic uncertainties of the $\phi^{\prime}$ parameters were analyzed:

\begin{itemize}
\item{
The systematic uncertainty of cross section measurement induces 5.1\% uncertainty of $\Gamma^{\phi^{\prime}}_{ee}\mathcal{B}^{\phi^{\prime}}_{\phi\eta}$ and $\mathcal{B}^{\phi^{\prime}}_{e^{+}e^{-}}\mathcal{B}^{\phi^{\prime}}_{\phi\eta}$.
}
\item{
The uncertainty of the branching fractions of $\phi^{\prime}$-meson decay channels causes the uncertainty of $\phi^{\prime}$ shape. According to~\cite{pdg19} the relative uncertainties of $\mathcal{B}^{\phi^{\prime}}_{K^{*}(892)K}$, $\mathcal{B}^{\phi^{\prime}}_{\phi\eta}$ and $\mathcal{B}^{\phi^{\prime}}_{\phi\sigma}$ can be estimated as $15\%$, $30\%$ and $15\%$, correspondingly. The variation of the branchings within these uncertainties with the requirement $\mathcal{B}^{\phi^{\prime}}_{K^{*}(892)K}+\mathcal{B}^{\phi^{\prime}}_{\phi\eta}+\mathcal{B}^{\phi^{\prime}}_{\phi\sigma} \equiv 1$ leads to the uncertainties of 3 eV for $\Gamma^{\phi^{\prime}}_{ee}\mathcal{B}^{\phi^{\prime}}_{\phi\eta}$, 4 MeV for $m_{\phi^{\prime}}$ and 13 MeV for $\Gamma_{\phi^{\prime}}$.
}
\item{
The contribution of the uncertainty of nonresonant amplitude energy dependence was studied by 
performing the fit with different non-$\phi^{\prime}$ amplitudes: 
$0$, $a_{\rm n.r.}$, $a_{\rm n.r.}/s^{3/2}$, $a_{\rm n.r.}/s$, $a_{\rm n.r.}/\sqrt{s}$, 
$a_{\rm n.r.}{\cdot}\sqrt{s}$, $a_{\rm n.r.}{\cdot}s$ ($a_{\rm n.r.}$ is constant). 
The resulting $\phi^{\prime}$ uncertainties are 14 eV for $\Gamma^{\phi^{\prime}}_{ee}\mathcal{B}^{\phi^{\prime}}_{\phi\eta}$, 10 MeV for $m_{\phi^{\prime}}$ and 36 MeV for $\Gamma_{\phi^{\prime}}$.
}
\end{itemize}

The overall systematic uncertainties of the $\phi^{\prime}$ parameters, shown in Table~\ref{fit_results}, are obtained by a quadratic summation of the listed individual uncertainties. 

\begin{table}[H]
  \begin{center}    
\caption{Systematic uncertainties of the $\sigma(e^{+}e^{-}{\to}\phi\eta)$ measurement. \label{tab:syst}}
    \begin{tabular}{lc}
      \hline
      Source & Value, \%\\ 
      \hline
      Event selection & 1.6 \\
      \hline
      Signal/background separation  & 4.1 \\
      \hline
      Efficiency correction & 2.1  \\     
      \hline
      Luminosity & 1 \\
      \hline
      $\mathcal{B}^{\phi}_{K^+K^-}$ & 1 \\
      \hline
      Overall & 5.1\\
      \hline      
    \end{tabular}
  \end{center}
\end{table}


\section{Contribution to $(g-2)_\mu$}
\hspace*{\parindent}
Using the result obtained for the $e^+e^- \to \phi\eta$ cross section we calculate the corresponding leading-order hadronic contribution to the anomalous magnetic moment of muon $a_{\mu}$. 
According to Ref.~\cite{g2} this contribution for the $E_{\rm c.m.}$ range from $E_{\rm min} \equiv 2m_{K^{+}}+m_{\eta}$ to $E_{\rm max}$ is expressed as

\begin{eqnarray}
  \label{gm2formula}
  a_{\mu}^{\phi\eta}(E<E_{\rm max}) = \Biggr(\frac{\alpha m_{\mu}}{3\pi}\Biggr)^{2} \int_{E^2_{\rm min}}^{E^2_{\rm max}} \frac{ds}{s^2} K(s)\cdot \frac{\sigma(e^+e^-\to \phi\eta)|1-\Pi(s)|^2}{\sigma_0(e^+e^-\to \mu^+\mu^-)}, 
\end{eqnarray}
where $K(s)$ is the kernel function, the factor $|1 - \Pi(s)|^2$ excludes 
the effect of leptonic and hadronic vacuum polarization (VP), and $\sigma_0(e^+e^-\to \mu^+\mu^-) = 4\pi\alpha^2/(3s)$. 
The integration is performed using the trapezoidal method and based on the experimental cross section values. The calculation of $a_{\mu}^{\phi\eta}$ for $E_{\rm max}=1.8$ and 2.0 GeV gives

\begin{eqnarray}
a_{\mu}^{\phi\eta}(E<1.8\, {\rm GeV})=(0.321 \pm 0.015_{\rm  stat} \pm 0.016_{\rm  syst}) \times 10^{-10},\nonumber \\
a_{\mu}^{\phi\eta}(E<2.0\, {\rm GeV})=(0.440 \pm 0.015_{\rm  stat} \pm 0.022_{\rm  syst}) \times 10^{-10}.\nonumber
\end{eqnarray}

Here the first uncertainty is statistical, the second one corresponds to the systematic uncertainty of $\sigma(e^+e^- \to \phi\eta)$. These values should be compared to the calculations, based on BaBar data in the corresponding $E_{\rm c.m.}$ ranges (see~\cite{davier,teubner}):

\begin{eqnarray}
a_{\mu}^{\phi\eta}(E<1.8\, {\rm GeV})=(0.36 \pm 0.02_{\rm  stat} \pm 0.02_{\rm  syst})\times 10^{-10},\nonumber \\
a_{\mu}^{\phi\eta}(E<2.0\, {\rm GeV})=(0.46 \pm 0.03_{\rm tot})\times 10^{-10}.\nonumber
\end{eqnarray}
Here for $a_{\mu}^{\phi\eta}(E<2.0\, {\rm GeV})$ the total uncertainty is shown. Note, that the work~\cite{davier} uses the quadratic interpolation between data points, while in the work~\cite{teubner} the trapezoidal rule is used. It is seen that our values for $a_{\mu}^{\phi\eta}$ for $E_{\rm c.m.}<1.8$ GeV are about $1\sigma$ lower than previous results. 


\section{Conclusion}
\hspace*{\parindent}
The process $e^+e^-{\to}K^+K^-\eta$ has been studied in the center-of-mass 
energy range from 1.59 to 2.01\,GeV using the data sample of 59.5 pb$^{-1}$ 
collected with the CMD-3 detector. In the production of the $K^+K^-\eta$ 
final state we observed the contribution of the $\phi(1020)\eta$ intermediate 
state only. On the base of 3009 $\pm$ 67 selected signal events the cross 
section of $e^+e^-{\to}\phi(1020)\eta$ has been measured with the systematic 
uncertainty of 5.1$\%$. The obtained cross section has been used to calculate 
the contribution to the anomalous magnetic moment of the muon:
$a_{\mu}^{\phi\eta}(E<1.8\, {\rm GeV})=(0.321 \pm 0.015_{\rm  stat} \pm 0.016_{\rm  syst}) \times 10^{-10}$, 
$a_{\mu}^{\phi\eta}(E<2.0\, {\rm GeV})=(0.440 \pm 0.015_{\rm  stat} \pm 0.022_{\rm  syst}) \times 10^{-10}$.
From the $e^+e^-{\to}\phi(1020)\eta$ cross section 
approximation the $\phi(1680)$ meson parameters have been determined with 
precision comparable or better than in previous measurements.

\section{Acknowledgment}
\hspace*{\parindent}
We thank the VEPP-2000 personnel for excellent machine operation. The work is partially supported by the Russian 
Foundation for Basic Research grants 17-52-50064-a, 17-02-00897. 
Part of this work related to simulation of multihadronic production is supported by the MSHE grant 14.W03.31.0026.



\begin{thebibliography}{00}

\bibitem{fred} F. Jegerlehner, Springer Tracks Mod. Phys. {\bf 274}, 1 (2017).

\bibitem{davier}  M. Davier, A. Hoecker, B. Malaescu, and Z. Zhang, Eur. Phys. J. C {\bf 77}, 827 (2017).

\bibitem{thomas} A. Keshavarzi, D. Nomura, T. Teubner, Phys. Rev. D {\bf 97}, 114025 (2018).

\bibitem{teubner}  K. Hagiwara {\em et al}., J. Phys. G {\bf 38}, 085003 (2011).

\bibitem{bnl} G.W. Bennett  {\em et al}.  (Muon g-2 Collaboration), Phys. Rev. D {\bf 73}, 072003 (2006).

\bibitem{babar_kpkmeta_2gamma} B. Aubert {\em et al}. (BaBar Collaboration), Phys. Rev. D {\bf 77}, 092002 (2008).

\bibitem{babar_kpkmeta_pippimpi0} B. Aubert {\em et al}. (BaBar Collaboration), Phys. Rev. D {\bf 76}, 092005 (2007).

\bibitem{snd_kpkmeta} M.N.~Achasov {\em et al}. (SND Collaboration), Phys. Atom. Nuclei {\bf 81}, 205 (2018).

\bibitem{vepp1} V.V.~Danilov {\em et al.}, Proceedings EPAC96, Barcelona, p.1593 (1996).

\bibitem{vepp2} I.A.~Koop, Nucl. Phys. B (Proc. Suppl.) {\bf 181-182}, 371 (2008).
  
\bibitem{vepp3} P.Yu.~Shatunov {\em et al.}, Phys. Part. Nucl. Lett. {\bf 13}, 995 (2016).

\bibitem{vepp4} D.~Shwartz {\em et al}., PoS ICHEP2016, 054 (2016).

\bibitem{cmd3} B.I. Khazin  {\em et al}.  (CMD-3 Collaboration), Nucl. Phys. B (Proc. Suppl.) {\bf 181-182}, 376 (2008).

\bibitem{dc} F.~Grancagnolo {\em et al.}, Nucl. Instr. Meth. A{\bf 623}, 114 (2010).

\bibitem{lxe} A.V.~Anisyonkov {\em et al.}, Nucl. Instr. Meth. A{\bf 598}, 266 (2009).

\bibitem{cal} D.~Epifanov (CMD-3 Collaboration), J.\ Phys.\ Conf.\ Ser.\  {\bf 293}, 012009 (2011).

\bibitem{GEANT4} S. Agostinelli  {\em et al}. (GEANT4 Collaboration), Nucl. Instr. and Meth. A {\bf 506}, 250 (2003).

\bibitem{lum} A.E. Ryzhenenkov  {\em et al}. (CMD-3 Collaboration), EPJ Web Conf., 212, 04011 (2019).

\bibitem{laser} E.V. Abakumova  {\em et al}., Phys. Rev. Lett. {\bf 110}, 140402 (2013).

\bibitem{laser2} E.V. Abakumova  {\em et al}., JINST {\bf 10}, T09001 (2015).

\bibitem{pp} R.R. Akhmetshin  {\em et al}.  (CMD-3 Collaboration), Phys. Lett. B {\bf 759}, 634 (2016).

\bibitem{shemyakin_kkpipi} D.N. Shemyakin {\em et al}. (CMD-3 Collaboration), Phys.Lett. B {\bf 756}, 153 (2016).

\bibitem{babar_kkpipi} J.P. Lees {\em et al}.  [BaBar Collaboration], Phys. Rev. D {\bf 86}, 012008 (2012).

\bibitem{kur_fad} E.A. Kuraev and V.S. Fadin, Sov. J. Nucl. Phys. {\bf 41}, 466 (1985).

\bibitem{OZI} S. Okubo, Phys. Lett, {\bf 5}, 165 (1963); G. Zweig, CERN report S419/TH412 (1964), unpublished; I. Iizuka, K. Okada, and O. Shito, Prog. Theor. Phys. {\bf 35}, 1061 (1966).

\bibitem{pdg19} M. Tanabashi {\em et al}. (Particle Data Group), Phys. Rev. D {\bf 98}, 030001 (2018) and 2019 update. 

\bibitem{g2} A. Hoefer, J. Gluza, F. Jegerlehner, Eur. Phys. J. C {\bf 24}, 51 (2002).

\end{thebibliography}
\end{document}